# Nonparametric Identification in Index Models of Link Formation

Wayne Yuan Gao*

May 15, 2018

**Abstract**

We consider an index model of dyadic link formation with a homophily effect index and a degree heterogeneity index. We provide nonparametric identification results in a single large network setting for the potentially nonparametric homophily effect function, the realizations of unobserved individual fixed effects and the unknown distribution of idiosyncratic pairwise shocks, up to normalization, for each possible true value of the unknown parameters. We propose a novel form of scale normalization on an arbitrary interquantile range, which is not only theoretically robust but also proves particularly convenient for the identification analysis, as quantiles provide direct linkages between the observable conditional probabilities and the unknown index values. We then use an inductive "in-fill and out-expansion" algorithm to establish our main results, and consider extensions to more general settings that allow nonseparable dependence between homophily and degree heterogeneity, as well as certain extents of network sparsity and weaker assumptions on the support of unobserved heterogeneity. As a byproduct, we also propose a concept called "modeling equivalence" as a refinement of "observational equivalence", and use it to provide a formal discussion about normalization, identification and their interplay with counterfactuals.

*Gao: Department of Economics, Yale University, 28 Hillhouse Ave, New Haven, CT 06511, USA; wayne.gao@yale.edu. I am grateful to Peter C.B. Phillips and Xiaohong Chen for their invaluable advice, guidance and encouragement, and I thank Donald Andrews, Tim Armstrong, Eric Auerbach, Aureo de Paula, Phil Haile, Sukjin Han, Yuichi Kitamura, Michael Leung, Rosa Matzkin and Peter Toth for helpful discussions, comments and suggestions.

# Contents

# 1 Introduction

Bilateral relationships ("links") are one of the most basic forms of social and economic relationships. For instance, trades are often carried out bilaterally between a seller and a buyer, contracts are often signed bilaterally by two counterparties, communication of information often happens bilaterally between a sender and a receiver, and friendship is often a bilateral relationship between two individuals. Such connections follow almost trivially from the observation that economics is largely concerned with individual decision making, in which an economic agent is the most basic unit of analysis, thus leading to bilateral relationships as the simplest form of relationships. Of course, in general equilibrium theory and game theory, multilateral relationships are also important, and yet many multilateral linkages are ultimately the outcome of a collection of bilateral relationships. The rapidly growing recent literature on the econometrics of networks may be viewed as an endeavor to model complex social and economic relationships explicitly in terms of a collective of bilateral linkages that interact in potentially sophisticated ways.

One line of research in network econometrics focuses on dyadic link formation, specifically on how individual characteristics affect the stochastic formation of bilateral relationships. A recent advance by Graham (2017) considers a link formation model with homophily effects and individual degree heterogeneity of the following form:

$$D_{ij} = \mathbf{1}\left\{W_{ij}'\beta_0 + A_i + A_j \geq U_{ij}\right\}, \quad U_{ij} \sim_{iid} Logit \tag{1}$$

This model has two systematic drivers of link formation. Specifically, the homophily effects index $W_{ij}'\beta_0$ captures the tendency of individuals to link ($D_{ij} = 1$) with individuals of similar *observable* (by the econometrician) characteristics,[1] such as ethnicity and educa-tion. The degree heterogeneity index $(A_i + A_j)$ captures variations generated by *unobserved* individual fixed effects $A_i$ and $A_j$, such as "emotional intelligence" or "coolness". As discussed in Graham (2017) and earlier research cited therein, the presence and magnitude of homophily effects and degree heterogeneity may have empirical and policy implications on multiple aspects of social networks, such as information diffusion, social learning, epidemic contagion and racial segregation. Within this framework, Graham provides consistent and asymptotically normal maximum-likelihood estimates for the homophily effect parameters $\beta_0$, assuming that the exogenous idiosyncratic pairwise shocks $U_{ij}$ are independently and identically distributed with a logistic distribution.

[1]More specifically, the observable pairwise variable $W_{ij} = \tilde{w}(X_i, X_j)$ is modeled as a known symmetric function $\tilde{w}(\cdot,\cdot)$ of the pair's individual observable characteristics $X_i$ and $X_j$.

This paper start by considering a generalized version of model (1) in the following form:

$$D_{ij} = \mathbf{1}\left\{w\left(X_i, X_j\right) + A_i + A_j \geq U_{ij}\right\}, \quad U_{ij} \sim_{iid} F, \tag{2}$$

where the homophily effect function $w$ and the cumulative distribution function $F$ are both taken to be *unknown*. We then establish that, under mild assumptions on the unknown functions and the data distribution, almost all unknown aspects of model (2) are identified up to normalization. Moreover, we provide an explicit characterization of the sharp identified set, an equivalence class induced by an exact form of positive affine transformations. We discuss why an explicit account of the invariant properties of the sharp identified set is crucial for interpreting the identification results in a meaningful way, and for determining the identifiability of counterfactual parameters as well as the answerability of economic questions about model (2).

Our main identification arguments start with a novel strategy to impose scale normalization, where we normalize as unity an arbitrarily chosen *interquantile range* (IQR)[2] of the CDF $F$. We formally show that the interquantile-range normalization proposed in this paper is guaranteed to be valid and without loss of generality, whenever $F$ is nondegenerate. This paper, according to our present knowledge, is one of the first in the econometric literature to apply the well-known idea of using interquantile ranges as a robust measure of dispersion to identification analysis in binary response models.

The interquantile-range normalization turns out particularly advantageous in our context, as it helps establishing a simple and direct linkage between the observable conditional linking probability and the unknown index values, which serves as a lever in our identification analysis. Specifically, the interquantile-range scale normalization, along with another location normalization on one of the quantiles, together normalize the locations of two different quantiles of $F$. Our main identification arguments then demonstrate how to use the two normalized quantiles to characterize the whole unknown $F$ in a convenient way. Specifically, we initially identify individual fixed effects that correspond to the two normalized quantiles, and then employ an inductive "in-fill" and "out-expansion" algorithm to recursively establish joint identification of the whole unknown CDF $F$ and all realized fixed effects $A_i$. Then, by considering individuals with arbitrary observable characteristics but particular fixed effects, we are able to identify the homophily effects simply by inverting the identified CDF.

[2] Note that the abbreviation "IQR" usually stands for the "*interquartile range*", defined as the difference between the third and the first quartiles of a given distribution. Clearly, the interquartile range is a particular interquantile range.

As extensions of the primary formulation (2), we also consider specifications in the following form,

$$D_{ij} = \mathbf{1}\left\{w\left(X_i, X_j\right) + \phi\left(A_i, A_j\right) \geq U_{ij}\right\}, \quad U_{ij} \sim_{iid} F,$$

where $\phi$ is allowed to be nonlinear in different manners, or to be even unknown. With a known nonlinear function $\phi$, we characterize general sufficient conditions in terms of homogeneity and translatability of the function $\phi$, under which the location or scale of $F$ becomes unidentifiable while the equivalence relation underlying the lack of identification becomes precisely characterizable. This is consistent with the well-known observation in the previous literature that a known form of nonlinearity helps with identification in general, and we show how the identification arguments for our main model (2) can be adapted to accommodate and exploit the known form of nonlinearity using a "conjecture and falsification" procedure. With $\phi$ taken to be unknown, we show that a further normalization can reduce $\phi$ to a simple linear sum when two individuals share the same level of fixed effects. Finally, we also investigate a specification under complete lack of additive separability between homophily effects and degree heterogeneity,

$$D_{ij} = \mathbf{1}\left\{\phi\left(X_i, X_j; A_i, A_j\right) \geq U_{ij}\right\}, \quad U_{ij} \sim_{iid} F,$$

where $\phi$ is again taken to be unknown. We show that the model admits even more flexible normalization and obtain a simpler characterization of the identification relationship.

Finally, we show that the identification arguments may be adapted to accommodate a certain extent of network sparsity, where linking probabilities converge to zero at a rate slower than $n^{-1}$. Moreover, we show (in the appendix) that the identification strategy continues to be applicable even when the support of individual degree heterogeneity is bounded and unknown.

This paper belongs to an extensive line of literature in statistics and econometrics that study stochastic network formation models. In statistics, the most related line of literature considers the so-called "$\beta$-model", which dates back to Bradley and Terry (1952), Holland and Leinhardt (1981) and Newman, Strogatz, and Watts (2001), and is further developed by Blitzstein and Diaconis (2011), Chatterjee, Diaconis, and Sly (2011), Yan and Xu (2013) and Yan, Leng, and Zhu (2016). These papers mainly consider parametric variants (primarily within the exponential family) of the main model (2), mostly without observable individual characteristics.

In econometrics, this paper belongs to the line of literature that studies dyadic link formation in a single large network setting. See Graham (2015), Chandrasekhar (2016) and de Paula (2016) for a review of other lines of literature on the econometrics of network models. Besides Graham (2017), to which this paper is most closely related, Charbonneau (2017) considers a setup similar to Graham (2017) in a panel data setting, while Dzemski (2017), Jochmans (2017) and Yan, Jiang, Fienberg, and Leng (2018) study maximum likelihood estimators of homophily effects in directed network formation. Also related are Candelaria (2016) and Toth (2017), who consider semiparametric identification and estimation of the link formation model under a linear specification of homophily effects, assuming that the homophily effect parameter is nonzero. We further consider the case of nonparametric homophily effects, which requires a very different type of identification strategy: instead of differencing out fixed effects, we start by first effectively "shutting down" the index of observable characteristics and directly confronting the identification of realized fixed effects and the unknown distribution of idiosyncratic pairwise shocks, after which the homophily effects become almost straightforward to identify. This approach to fixed effects is in spirits closely related to the recent work by Bonhomme, Lamadon, and Manresa (2017)who propose a two-step estimation procedure in a different panel data setting that utilizes the *kmeans* algorithm in the first step to estimate discretized fixed effects and then plug them into a likelihood-based second step to parametrically estimate the structural parameter.

Relatedly, this paper also contributes to a line of research that utilizes the form of link formation models considered here in order to study structural social interaction models: for instance, Arduini, Patacchini, and Rainone (2015), Auerbach (2016), Goldsmith-Pinkham and Imbens (2013), Hsieh and Lee (2016) and Johnsson and Moon (2017). In these papers, the social interaction models are the main focus of identification and estimation, while the link formation models are used mainly as a tool (a control function) to deal with network endogeneity or unobserved heterogeneity problems in the social interaction model. As the link formation models are merely means to an end, this line of literature is usually not concerned with full identification of the link formation model. On one hand, the observation that the link formation model considered here is used as a workhorse model in these studies also motivates our current endeavor to investigate its nonparametric identifiability. On the other hand, the full identification results established here for the link formation model, especially the realized fixed effects and the unknown CDF, are suited for a different set of counterfactual analysis: while the social interaction models may be useful for evaluating how policy intervention in network structure may affect certain outcomes via social influence, our link formation models are how policy intervention in certain dimensions of observable characteristics may induce endogenous changes in network formation.

It should be pointed out that in this paper we do not consider link externality in network formation, as studied in a series of other papers such as De Paula, Richards-Shubik, and Tamer (2018), Graham (2016), Leung (2015), Menzel (2015), Mele (2017a), Mele (2017b) , Ridder and Sheng (2017) and Sheng (2016). However, the focus and the methodology of these papers are noticeably different from ours. Moreover, as shown in Mele (2017a), with homogeneous agents and nonnegative link externality, a class of exponential random graphs (ERGM) is asymptotically indistinguishable from the Erdos-Renyi random graph models with independent links under suitable conditions. This heuristically suggests that dyadic network formation models may be regarded as a reduced-form approximation of the steady-states of more sophisticated network formation models with nonnegative link externality.

Lastly, this paper is also related to the general literature on nonparametric and semiparametric identification, normalization and their interplay with counterfactual analysis. Given the huge amount of work that dates back to more a century ago and involves an extensive variety of concepts, we refrain from reviewing the related literature here. See the survey articles by Matzkin (2007) and more generally by Lewbel (2018) for a discussion of these topics. As a by-product (presented mainly in Section 3 and Appendix B), this paper proposes a potentially complementary approach to the formalization of concepts such as *modeling equivalence* (a proposed refinement of "observational equivalence"), normalization and identifiability of counterfactuals via an abstract argument based on bijective mappings.

The rest of the paper is organized as follows. In Section 2, we describe the main specifications of the link formation model, lay out the corresponding assumptions and present our identification results. In Section 3, we provide a discussion about normalization without loss of generality, in view of its importance for the interpretation of our identification results and arguments. Section 4 contains the main proof for our identification results. Section 5 considers extensions of the main model to accommodate nonlinear or unknown coupling of individual fixed effects, nonseparable dependence between homophily effects and fixed effects, as well as a certain extent of network sparsity. Some conclusions and further research directions are discussed in Section 6. Further discussions of assumptions, extensions, and technical details are available in the Appendix.

# 2 Basic Model and Main Results

## 2.1 Model Specification

In Sections 2-4, we consider the dyadic link formation model (3) below with potentially nonparametric homophily effect and linearly additive individual degree heterogeneity:

$$D_{ij} = \mathbf{1}\left\{w\left(X_i, X_j\right) + A_i + A_j \geq U_{ij}\right\}, \quad \forall i \neq j \in N, \tag{3}$$

where :

- $N = \{1, ..., n\}$ is a set of $n$ individuals.
- $D_{ij}$ is an observable symmetric binary indicator variable that represents whether there is a link between $ij$ ($D_{ij} \equiv D_{ji} = 1$) or not ($D_{ij} \equiv D_{ji} = 0$).
- $X_i$ is an observable $k$-dimensional random vector of characteristics of individual $i$ with support $Supp\left(X_i\right) \subseteq \mathbb{R}^k$.
- $w : Supp\left(X_i\right) \times Supp\left(X_i\right) \to \mathbb{R}$ is an *unknown* symmetric function that produces an *unobservable* dyad-level index

  $$W_{ij} \equiv W_{ji} := w\left(X_i, X_j\right) \equiv w\left(X_j, X_i\right) \in \mathbb{R},$$

  which we intend to interpret as the homophily effects:
- $A_i$ is an *unobservable* scalar random variable that captures individual degree heterogeneity, which summarizes how an individual $i$'s unknown characteristics contributes to this person's popularity via the individual fixed effect on the value of linkage.
- $U_{ij} \equiv U_{ji}$ is an *unobservable* scalar random variable that captures the dyad-level idiosyncratic shock to the value of link $ij$, whose distribution is *unknown*.

Our model (3) is a natural generalization of the ones considered in Graham (2017), Candelaria (2016) and Toth (2017), all of which impose parametric assumptions on the homophily effect index. Formally, they specify

$$w\left(X_i, X_j\right) := \tilde{w}\left(X_i, X_j\right)' \beta_0 \tag{4}$$

where $\beta_0 \in \mathbb{R}^k$ is an *unknown* finite-dimensional parameter and $\tilde{w} : \mathbb{R}^k \times \mathbb{R}^k \to \mathbb{R}^k$ is a *known* function that produces an observable $k$-dimensional pairwise characteristics ("distances") $\tilde{w}(x_i, x_j)$. In view of the popularity of this parametric specification and its relevance in applied work, we also present specific results related to this parametric specification, and throughout the paper we use the tilde notations "$\tilde{w}$" as in (4) to indicate this parametric specification.

From now on, we refer to model (3) with nonparametric homophily effects as the "*nonparametric homophily setting*", and refer to model (3) under (4) with parametric homophily effects as the "*parametric homophily setting*".

## 2.2 Nonparametric Homophily Setting

We first focus on the nonparametric homophily setting in this subsection, and present the corresponding assumptions along with a discussion, followed by a formal statement of the identification result.

**Assumption 1** (Degeneration of Homophily Function on the Hyper-Diagonal)**.** *The homophily effect function $w : Supp(X_i) \times Supp(X_i) \to \mathbb{R}$ is symmetric in its two vector arguments, continuous with respect to the Euclidean norm on $Supp(X_i) \times Supp(X_i) \subseteq \mathbb{R}^{2k}$, and takes an unknown constant value $\theta \in \mathbb{R}$ on the hyper-diagonal of $\mathbb{R}^k \times \mathbb{R}^k$:*

$$\left\{(x_i, x_j) \in \mathbb{R}^k \times \mathbb{R}^k : x_i = x_j\right\}.$$

**Assumption 2** (Exogenous IID $U_{ij}$ with Nice CDF)**.** *$U_{ij} \perp (X, A)$ for all $ij$, and $\{U_{ij}\} \sim_{iid} F$, where $F$ is an unknown strictly increasing and continuous CDF on $\mathbb{R}$.*

**Assumption 3** (Random Sampling)**.** *$(X_i, A_i)$ are i.i.d. across $i \in N$.*

**Assumption 4** (Conditional Full Support of $A_i$)**.** *$\forall x \in Supp(X_i)$, $Supp(A_i | X_i = x) \equiv \mathbb{R}$.*

**Discussion of Assumptions**

Assumption 1 is key for the identification arguments in this paper: it basically allows us to consider pairs of individuals with observable characteristics that are known to produce some degenerate homophily effects. Specifically, even though $w : Supp(X_i) \times Supp(X_i) \to \mathbb{R}$ is an unknown function, Assumption 1 requires that a "level curve" of $w$ is known ex ante to be the hyper-diagonal of $\mathbb{R}^k \times \mathbb{R}^k$, which is consistent with the interpretation of $w(X_i, X_j)$ as homophily effects that become degenerate when the observable characteristics of the two individuals coincide completely. First, notice that we do *not* require that, outside the hyper-diagonal, the homophily function $w$ take values different from $\theta$; in other words, we do not require that the known level curve coincide exactly with the whole level set. Second, Assumption 1 is weaker than the requirement that the homophily function be a (pseudo-)metric[3] on the space of observable characteristics, in which case the interpretation of $w(X_i, X_j)$ as "distance-based homophily effects" will become even clearer. See Assumption 1a, a parametric variation of Assumption 1, in Subsection 2.3 along with a discussion of its relationship to the corresponding assumptions made in Graham (2017), Candelaria (2016) and Toth (2017) under the parametric homophily setting. See Assumption 1' in Appendix A for a slightly weaker version that captures the essential requirement of Assumption 1. See Remark 3 in Appendix C.2 for a discussion on why Assumption 1 is no longer required if we consider certain nonlinear forms of fixed-effect coupling functions.

Assumption 2 states that the idiosyncratic pairwise shocks $U_{ij}$ are exogenously, independently and identically distributed with an unknown CDF with nice properties. It is similar to the corresponding assumptions in Candelaria (2016) and Toth (2017), where no parametric restriction is imposed on the distribution of the unobservable idiosyncratic pairwise shock $U_{ij}$. This paper, along with Candelaria (2016) and Toth (2017), relaxes the parametric assumption of logistic distribution in Graham (2017). Note that, even though we assume $F$ to be continuous, our main identification arguments can be adapted to accommodate jumps in $F$.[4] However, as this extension is rather trivial and technical, we focus on the case where $F$ is continuous.

Assumption 3 is a standard assumption about random sampling, which is essentially the same as those made in Graham (2017), Candelaria (2016) and Toth (2017).

Assumption 4 imposes a conditional full support condition for the individual fixed effects, which essentially requires that, for a given pair of individuals, regardless of the values of their observable characteristics, each of their unobserved individual heterogeneity can be potentially pivotal in determining their linkage or lack thereof. Assumption 4, as presented here, is not necessary: Proposition 4 in Appendix F shows that our identification strategy is adaptable to the case where the support of fixed effects is bounded. Assumption 4" in Appendix F is essentially equivalent to the corresponding assumption made in Graham (2017, Assumption 5), who assumes compact support of individual fixed effects $A_i$ in $\mathbb{R}$. Assumptions 4 or 4" together are essentially equivalent to the assumption made in Toth (2017), who requires $A_i|_{X_i=x}$ has a common support, which may or may not be the whole real line.

[3]For $w$ to be a (pseudo-)metric, it needs to satisfy triangular inequalities of the form $w(x_i, x_h) \leq w(x_i, x_j) + w(x_j, x_h)$ for all $x_i, x_j, x_h \in \mathbb{R}^k$, which we do not need for our results.

[4]See footnote 9.

With Assumptions 1, 2, 3 and 4, we now present our main result on identification under the nonparametric homophily setting:

**Theorem 1** (**Identification up to Normalization under Nonparametric Homophily**). *For the nonparametric homophily setting, under Assumptions 1, 2, 3 and 4,* $(w, A, F^{-1})$ *is identified up to location and scale normalization. More precisely, the set generated by all positive affine transformations of the true* $(\omega, A, F^{-1})$ *in the following form*

$$\left\{ \left( cw\left(\cdot,\cdot\right)+b,\ cA+a,\ cF^{-1}\left(\cdot\right)+2a+b \right) :\ a,b\in\mathbb{R}, c>0 \right\}$$

*is identified, and no proper nonempty subset of it can be further identified.*

The proof of Theorem 1 is presented in Sections 3 and 4.

Theorem 1 establishes that the class of link formation models in the form of (3) are (semi-)nonparametrically identified. The identification results are presented as sharp set identification, where the identified set is an equivalence class of the unknown parameters induced by a particular form of positive affine transformations that the true values belong to. This on one hand reflects the ultimate lack of point identification that is familiar in binary responses models, but on the other hand establishes that all points in the identified set are equivalent in a precise sense. Our identification result is also sharp: subsets of the identified equivalence classes are necessarily unidentified.

We refrain from a popular practice in the literature to present identification results as point identification after imposing a *particular* normalization. This approach is innocuous for the identification of the parameters in question per se, but becomes potentially confusing when it comes to the identification of counterfactual parameters, formulated as functions of the original parameters, as the structure of the identified equivalence classes will be made implicit after imposing a particular normalization.

To be more precise, the positive affine transformations that define the identified equivalence class above involve simultaneous scaling and translation transformations on $(w, A, F^{-1})$ in a particular form, preserving some invariant properties in terms of topological structures (openness), order-theoretical structures (orders) and geometric structures (shapes). These invariant properties of the identified equivalence class is fundamental to the identifiability of counterfactuals: only counterfactual parameters (or economic questions) that are defined as functions of these *invariant* properties may be regarded as point identified. See Lemma 4 and Theorem 4 in Appendix B for more formal and detailed discussions.

## 2.3 Parametric Homophily Setting

In this Subsection, we specialize the model (3) to the parametric homophily setting (4).

Assumptions 2, 3 and 4 from Subsection 2.2 are maintained under the parametric homophily setting as well. However, Assumption 1 needs to be adapted as Assumption 1a, and Assumption 5a needs to be imposed as an additional assumption.

**Assumption. 1a**(Degeneration of Parametric Homophily Effect on the Hyper-Diagonal). *Each component* $\tilde{w}_m : \mathbb{R}^k \times \mathbb{R}^k \to \mathbb{R}$ *of the known function* $\tilde{w} = (w_m)_{m=1}^k$ *is symmetric, continuous with respect to the Euclidean norm on* $\mathbb{R}^k$*, and* $\tilde{w}_m (x_i, x_j) = \mathbf{0} \in \mathbb{R}^k$ *whenever* $x_i = x_j$.

**Assumption. 5a**(Full-Dimensional Support of $\tilde{W}_{ij}$): *Supp* $\left(\tilde{W}_{ij}\right) =$ *Supp* $(\tilde{w}(X_i, X_j))$ *contains* $k$ *linearly independent nonzero vectors* $\left\{\overline{W}_m \in \mathbb{R}^k : m = 1, ..., k\right\}$.

### Discussion of Assumptions

Assumption 1a captures essentially the same idea as Assumption 1, but is adapted to the parametric homophily setting, where $w(x_i, x_j) = \tilde{w}(x_i, x_j)' \beta_0 \equiv \tilde{W}_{ij}' \beta_0$ and $\tilde{w}, \tilde{W}$ are either known or observable. In this case, $w(x_i, x_j) = 0$ whenever $\tilde{W}_{ij} = \mathbf{0}$, so what we need essentially is that, for any value $x_i$ in the support of observable characteristics, we can find some other value $x_j$ of individual observable characteristics such that $\tilde{W}_{ij} = \tilde{w}(x_i, x_j) = \mathbf{0}$. Assumption 1a imposes a stronger version of this requirement: $\tilde{w}(x_i, x_j) = \mathbf{0}$ whenever $x_i = x_j$, i.e., $\tilde{w}$ degenerates to the zero vector on the hyper-diagonal, again leading to a clearer interpretation of homophily effect based on similarity of observable characteristics. Assumption 1a is similar to the corresponding assumptions in Toth (2017), while Graham (2017) and Candelaria (2016), stating their specifications mainly in terms of $\tilde{W}_{ij} \equiv \tilde{w}(X_i, X_j)$ rather than $X_i$, do not particularly require that the zero vector lie in the support of $\tilde{W}_{ij}$. Yet homophily effects based on Euclidean distances are discussed in their work as a leading example for applications, which automatically satisfy Assumption 1a. We discuss in Appendix A more weakened or strengthened variations of Assumptions 1a.

Assumptions 5a is a full-dimensionality condition on the support of observable pairwise characteristics $\tilde{W}_{ij} \equiv \tilde{w}(X_i, X_j)$, which is only relevant when the homophily effect is parametrically and linearly specified: if *Supp* $\left(\tilde{W}_{ij}\right)$ is contained in a proper linear subspace of $\mathbb{R}^k$, then it is impossible to point identify the $k$-dimensional parameter $\beta_0 \in \mathbb{R}^k$. Assumption 5a, as in Graham (2017, Assumption 2), directly imposes a restriction on the support of $W_{ij}$. However, Assumption 5a is significantly different from that made in Graham (2017), who only requires that *Supp* $(W_{ij})$ be a compact subset of $\mathbb{R}^k$. We do not require compactness, but require full-dimensionality. Candelaria (2016, Assumption A2) states the full-dimensionality condition in terms of the differences of pairwise characteristics of the form $\tilde{W}_{ij} - \tilde{W}_{ik}$, rather than $\tilde{W}_{ij}$ directly, which serves a similar purpose. Appendix A also provides an alternative version of Assumption 5a in terms of $X_i$ directly, corresponding to the "nonempty interior" assumption in Toth (2017) that guarantees full dimensionality.

With Assumptions 1a and 5a above, along with Assumptions 2, 3 and 4 in the previous subsection, we are now ready to present the identification result under the parametric homophily setting:

**Theorem 2** (**Identification up to Normalization under Parametric Homophily**). *For the parametric homophily setting, under Assumptions 1a, 2, 3, 4 and 5a, $(\beta_0, A, F^{-1})$ is identified up to location and scale normalization: more precisely, the set generated by all positive affine transformations of the true $(\beta_0, A, F^{-1})$ in the following form*

$$\left\{\left(c\beta_0,\ cA + a,\ cF^{-1}(\cdot) + 2a\right) :\ a \in \mathbb{R}, c > 0\right\}$$

*is identified, and no proper nonempty subset of it can be further identified.*

The proof of Theorem 2 is presented in Subsection 4.5, based on an adaption of the proof under the nonparametric homophily setting. It should be pointed out that identification under the nonparametric homophily setting does not automatically imply identification of $\beta_0$ under the parametric homophily setting. In particular, Assumption 5a (full dimensionality) is required in addition.

# 3 Normalization

Recall that the link formation model is given by

$$D_{ij} = \mathbf{1}\left\{w(X_i, X_j) + A_i + A_j \geq U_{ij}\right\},$$

where only $(D_{ij}, X_i, X_j)$ are observable. Given the discrete nature of the indicator function, it is well-known that the unknowns cannot be point identified. To establish Theorems 1 and 2, we first impose some normalization that helps with the characterization and the proof of the identification results.

In this section, we impose three forms of normalization, two of which are location normalization, with the third being scale normalization. The scale normalization we propose is novel relative to the usual practice in the econometric literature, but is also simple and intuitive: we normalize an arbitrarily chosen *interquantile range* to be unity, based on the idea that interquantile ranges are robust measures of dispersion and scale. We further provide a discussion on the advantages of the interquantile-range normalization.

In Appendix B, we show why the proposed normalization is *without loss of generality* in the following precise sense: both the identifiability of any counterfactual parameters and the answerability of any economic questions about the unknown aspects of the model are invariant under the proposed normalization (Theorem 4). We define and analyze a series of concepts related to the "modeling equivalence transformation" that underlie the normalization we impose, leading to a partition of the space of the unknown into "modeling equivalence classes". We explain later the difference between the nonstandard terminology "*modeling equivalence*" we adopt in this paper and the widely accepted terminology "*observational equivalence*". We then provide a rigorous proof of "normalization without loss of generality", and discuss how these proposed concepts can be potentially useful for clarifying the subtlety involved in identification, normalization and counterfactual analysis. However, as this is tangential to the main topic of this paper, we defer the analysis and discussion to Appendix B.

We now lay out the normalization we impose. Specifically, for any scalar $a, b \in \mathbb{R}$, and any strictly positive scalar $c \in \mathbb{R}_{++} := (0, \infty)$, define for all $i, j$,

$$\begin{aligned} \hat{A}_i &:= cA_i + a, \\ \hat{w}(x_i, x_j) &:= cw(x_i, x_j) + b, \\ \hat{U}_{ij} &:= cU_{ij} + 2a + b. \end{aligned} \tag{5}$$

This essentially defines a transformation $\psi_{a,b,c}$ on the space of the unknown that maps each point $(w, A, U)$ to the point $\left(\hat{w}, \hat{A}, \hat{U}\right)$. First, the transformation $\psi_{a,b,c}$ is invertible (so that $\psi_{a,b,c}^{-1}$ is well-defined), surjective (so that all points are covered) and positively affine (so that orders and shapes are preserved); moreover, the inverse $\psi_{a,b,c}^{-1}$ is also a valid positive affine transformation in the form of (5).[5] Second, it is straightforward to check that, Assumption 1, 2, 3 and 4 are satisfied under $(D, X, w, A, U)$ if and only if they are satisfied under $\left(D, X, \hat{w}, \hat{A}, \hat{U}\right)$,[6] so that the imposed assumptions are invariant under the transformation (5). Third, the link formation equation (3) also remains invariant:

$$\hat{w}(X_i, X_j) + \hat{A}_i + \hat{A}_j \geq \hat{U}_{ij} \quad \Leftrightarrow \quad w(X_i, X_j) + A_i + A_j \geq U_{ij}.$$

[5]Formally, $\psi_{a,b,c}^{-1} = \psi_{-a,-b,c^{-1}} = \psi_{\hat{a},\hat{b},\hat{c}}$, where $\hat{a}, \hat{b} \in \mathbb{R}, \hat{c} \in \mathbb{R}_{++}$.

[6]Of course, some trivial corresponding adaptions in notations are needed: for example, the unknown CDF of the idiosyncratic pairwise shocks should now be written as "$\hat{F}$".

In other words, $\left(D, X, \hat{w}, \hat{A}, \hat{U}\right)$ are "observationally equivalent"[7] to $(D, X, \omega, A, U)$ . In view of the invariance of all modeling specifications and assumptions under the transformation $\psi_{a,b,c}$ defined in (5), we call call $\psi_{a,b,c}$ a *modeling equivalent transformation* of the unknown $(w, A, U)$ of our model. For the formal definitions, see Definitions 5-2 in Appendix B.

We now provide a discussion about the difference between the concept of *"modeling equivalence"* adopted here and the concept of *"observational equivalence*" widely used in the literature. In short, a transformation (via reparametrization or reformulation) of the model is regarded "*observationally equivalent*" if equivalence of the models with respect to the *observable data distribution* is preserved. In comparison, a transformation of the model is defined as "*modeling equivalent*" if equivalence of the models with respect to *both* the observable data distribution *and a set of maintained assumptions* is preserved. Hence, "modeling equivalence" should be regarded as a refinement of "observational equivalence" that restricts the flexibility of transformation operations along certain dimensions.

This, however, is best illustrated with a simple example that maintains observational equivalence but not modeling equivalence.

**Example 1** (Observational equivalence does *not* imply modeling equivalence.)**.** Suppose we redefine $\hat{A}_i := A_i - A_1$, and $\hat{U}_{ij} := U_{ij} - 2A_1$, then $\left(D, X, \hat{w}, \hat{A}, \hat{U}\right)$ are still *observationally equivalent* to $(D, X, \omega, A, U)$, as the transformation results in no changes whatsoever in the distribution of observable data. However, the transformation from $(D, X, \omega, A, U)$ to $\left(D, X, \hat{w}, \hat{A}, \hat{U}\right)$ induces a violation of Assumption 3 : $A_i$ is i.i.d. but $\hat{A}_i$ is clearly no longer i.i.d, and thus does *not* preserve *modeling equivalence* with respect to both the data distribution and the set of maintained assumptions (to which Assumption 3 belong). We might as well call the transformed model $\left(D, X, \hat{w}, \hat{A}, \hat{U}\right)$ together with a corresponding *simultaneous* transformation of Assumption 3 an "observationally equivalent model *reformulation*", the analysis of is beyond the intended scope of this paper.

Given that there are infinitely many possible ways to reformulate any econometric model without loss of observational equivalence, we argue that, by focusing on modeling equivalence, which is anchored not only by the observable data distribution but also by a set of *maintained assumptions*, we may pre-impose certain preferred structures that are motivated by sensible economic and statistical principles, and forbid any observationally equivalent transformations that violate such assumptions. For example, we may have good reasons to insist on assumptions of very basic topological structures such as continuity and connected-ness of certain spaces, and may not be interested in an observation equivalence class that does not respect such basic topological assumptions.

[7] "Observational equivalence" is usually defined by the invariance of the distribution of observable data under a transformation of the model. For a formal version of definition, see Appendix B.

In view of ((5)), we have two degrees of freedom (in terms of $a, b$) in location normalization and one degree of freedom (in terms of $c$) in scale normalization. We impose the first location normalization by setting

$$\theta \equiv w\left(x_i, x_i\right) = 0, \tag{6}$$

which is aligned with the interpretation of $w$ as a homophily effect function.

Second, we impose another location normalization, on the distribution of the random variable $U_{ij}$. A popular location normalization of this sort in the literature is to set $\mathbb{E}\left[U_{ij}\right] = 0$, provided that $\mathbb{E}\left[U_{ij}\right]$ exists, but in this paper we adopt a another popular approach that imposes normalization on a certain quantile, dating back to Manski (1985). Specifically, fixing some $\alpha \in \left(0, \frac{1}{2}\right)$, we normalize the $\alpha$-th quantile of the distribution of $U_{ij}$ to be zero:

$$Q_\alpha\left[U_{ij}\right] = 0. \tag{7}$$

The $\alpha$-th quantile of any almost-surely finite distribution is guaranteed to exist. Under Assumption 2, $F$ is strictly increasing, so $Q_\alpha\left[U_{ij}\right] = F^{-1}(\alpha)$ is well defined in $\mathbb{R}$, and the normalization of $\alpha$-th quantile is then equivalent to a normalization of the well-defined function $F^{-1}$ by setting $F^{-1}(\alpha) = 0$.

Third, we impose a scale normalization, also on the distribution of the random variable $U_{ij}$. Recall that we have normalized $Q_\alpha\left[U_{ij}\right] = 0$ in the second location normalization above for some fixed $\alpha \in \left(0, \frac{1}{2}\right)$. Under Assumption 2, $Q_{1-\alpha}\left[U_{ij}\right] - Q_\alpha\left[U_{ij}\right] > 0$ for any $\alpha \in \left(0, \frac{1}{2}\right)$, so we may normalize the "interquantile range", the difference between these two quantiles, to unity:

$$Q_{1-\alpha}\left[U_{ij}\right] - Q_\alpha\left[U_{ij}\right] = 1. \tag{8}$$

Equivalently, we normalize $F^{-1}(1-\alpha) - F^{-1}(\alpha) = 1$. As scale normalization on interquantile range seems relatively new to the identification literature, we now provide a brief discussion of its several appealing features.

It should be emphasized that the selection of the two quantiles $Q_\alpha\left[U_{ij}\right]$ and $Q_{1-\alpha}\left[U_{ij}\right]$ for location and scale normalization is merely one convenient choice among infinitely many alternatives. In the current context, we could normalize $Q_\alpha\left[U_{ij}\right], Q_\beta\left[U_{ij}\right] - Q_\alpha\left[U_{ij}\right]$ for any arbitrary $\alpha, \beta \in (0,1)$ with $\alpha < \beta$. We take $\alpha \in \left(0, \frac{1}{2}\right)$ and $\beta = 1 - \alpha$ simply to avoid unnecessary symbols.

Moreover, even though we set $Q_\alpha[U_{ij}], Q_{1-\alpha}[U_{ij}]$ to be 0 and 1, we refrain from attaching any substantial meaning to this "0" and this "1". This does not impose any *restriction* on the location and scale of the true underlying distribution. Instead, we should interpret any results, represented as functions of these normalized "0" and "1", as functions of $Q_\alpha[U_{ij}]$ and $Q_{1-\alpha}[U_{ij}]$, which intrinsically correspond to the probabilities $\alpha$ and $1-\alpha$.

The interquantile-range scale normalization is conceptually more general: for any non-degenerate distribution, there must exist two quantiles with well-defined strictly positive difference, which qualifies automatically as a proper divisor. A popular form of scale normalization is to set $Var[U_{ij}] = 1$ provided that $Var[U_{ij}]$ exists and is finite, which may not be true for heavy-tail distributions such as Cauchy distributions. Another approach of scale normalization, as proposed in Manski (1975, 1985, 1987) and also adopted in Candelaria (2016) and Toth (2017), involves setting the norm of some unknown parameters to unity, say, in our current context by setting $\|\beta_0\| = 1$ or $|\beta_{0,m}| = 1$ for some $m = 1, ..., k$ under the parametric homophily setting. This clearly rules a priori the case of $\beta_0 = \mathbf{0}$, which results in a strict loss of generality, and furthermore it induces a drastic change in basic topological structures of the parameter space. In particular, the normalization does not preserve openness, compactness and connectedness (if we augment the unit sphere with the origin or consider removal of a finite number of points), which may interfere with hypothesis testing of joint statistic significance. Technically, as we require $c \in (0, \infty)$ for $\psi_{a,b,c}$ defined in (5) to maintain equivalence in model specification and assumption, we cannot take $c = \sqrt{Var[U_{ij}]}$ or $c = \|\beta\|$, if $Var[U_{ij}]$ is not well-defined (or is $+\infty$) or $\|\beta\| = 0$. The interquantile-range normalization defined above, however, is not subject to these problems, thus allowing weaker conditions on the distribution of $U_{ij}$ or the structural parameters $\beta$.

Also, scale normalization on interquantile range is simple and intuitive. For a one-dimensional distribution, interquantile range normalization directly controls the length between two points, which is arguably one of the simplest quantity related to dispersion. The familiar concept of *interquartile* range, a particular interquantile range, has already been widely used as one of the most basic yet robust measure of dispersion, so it is also intuitive that interquantile ranges may used to control scales in identification analysis. In practice, it might be preferable to simply set $\alpha = 0.25$, so that $Q_{0.75}[U_{ij}] - Q_{0.25}[U_{ij}]$ corresponds exactly to the *interquartile* range, so that all *relative scale* identified subsequently may be interpreted with respect to the dispersion of random idiosyncratic errors as measured by the interquartile range.

Moreover, the interquantile-range scale normalization ($Q_{1-\alpha}[U_{ij}] - Q_\alpha[U_{ij}] = 1$) along with the quantile location normalization ($Q_\alpha[U_{ij}] = 0$), turns out to be particularly ad-vantageous in our current model, as it provides a direct and transparent linkage between conditional probabilities and the index values at two normalized quantiles $(Q_\alpha, Q_{1-\alpha})$. As conditional probabilities, with proper conditioning values, may be observed from data as frequencies, this linkage serves as a lever to convert knowledge from data into knowledge of index values. The exact way in which the two-quantile normalization works for our identification results is presented in Section 4. Yet the intuition above suggests that the interquantile-range normalization may be useful in other binary response models as well.

It is now worth pointing out that even though we impose the normalization above, the econometric *interpretation* of the model and the identification result should recognize the fundamental arbitrariness of this normalization. In the end, as explicitly stated in Theorems 1 and 2, we "*give back*" the particular normalization on the two quantiles imposed in this section, and present the final identification result in terms of sharp identified sets that are characterized by some precise forms of positive affine transformations. As has been discussed earlier, the transparency in the internal structures of the identified set makes explicit its invariant properties, including topological, order-theoretical and some precise geometric structures. In particular, the "shape" of the CDF $F$ may be regarded as point identified. Furthermore, the identifiability of counterfactual parameters is fully determined by invariance, or lack thereof, within the identified equivalence class.

It should then be further pointed out that the identification arguments to be presented in Section 4 can be carried out under $Q_\alpha [U_{ij}] = a,\ Q_\beta [U_{ij}] = b$ even without normalization, i.e., even if we leave $a$ and $b$ as unknown. This indicates that the two-quantile normalization consists of two conceptually different components: first, it is a normalization by our definition that labels modeling equivalent classes in a particular way; second, it is a method to characterize (the shape of) a CDF with respect to two quantiles of the CDF. The former is completely standard and done for the purpose of simpler notations, while the latter is an innovation that leverages the peculiar structures of binary response models. See Section C for further illustration on how the two-quantile characterization remains helpful even when we cannot make corresponding normalization.

Normalization, as defined here, is induced by the modeling equivalent transformation $\psi_{a,b,c}$ defined in (5), which is determined by all specifications and assumptions of our econometric model. Hence, any change in any specification or assumption may affect the validity of the normalization imposed above. The following remarks shortly discuss how the parametric specification of homophily (4) and the nonlinear specification of fixed effects affect the normalization above.

*Remark* 1 (**Parametric Specification of Homophily**). Under the parametric specification of homophily (4), the homophily effect function

$$w(x_i, x_j) = \tilde{w}(x_i, x_j)' \beta_0$$

is known up to a finite-dimensional parameter $\beta_0 \in \mathbb{R}^k$. In this case, the first location normalization becomes infeasible and unnecessary: $w(x_i, x_i) = 0$ is known for whatever true value of $\beta_0$. Hence, only the second location normalization (7) and the scale normalization will be needed. See Subsection 4.5 for more details.

*Remark* 2 (**Nonlinear Coupling of Fixed Effects**). In model (3), $A_i$ and $A_j$ are coupled into $A_i + A_j$ in a bilinear and additive way. Suppose we would like to consider the following alternative specification:

$$D_{ij} = \{w(X_i, X_j) + \phi(A_i, A_j) \geq U_{ij}\} \tag{9}$$

where $\phi : \mathbb{R}^2 \to \mathbb{R}$ is a *known*, nonlinear, symmetric, continuous, strictly increasing and surjective function. This immediately renders the second location normalization and the scale normalization infeasible: transforming $A_i$ to $cA_i + a$ results in a nonlinear change from $\phi(A_i, A_j)$ to $\phi(cA_i + a, cA_j + a)$, which may not maintain modeling equivalence to the original model. However, the proposed identification strategy based on the two-quantile characterization can be adapted. For example, if $\phi$ is a homogeneous function, say, $\phi(a_1, a_2) = (a_1 + a_2)^3$, then scale normalization is still feasible. In that case, even though location normalization is no longer feasible, the curvature of $\phi$ can be used to identify the location. See Subsection 5.1 and Appendix C for more details.

# 4 Proof of Main Results

We now present the proof for the identification results in Theorems 1 and 2.

By Lemma 2 in Appendix A, Assumptions 1a, 2, 3, 4 and 5a (for the parametric homophily setting) imply Assumption 1, 2, 3 and 4 (for the nonparametric homophily setting). Hence, we present the main identification arguments under Assumptions 1, 2, 3 and 4 for the nonparametric homophily setting in Subsections 4.1-4.4. In Subsection 4.5, we show that, with parametric specification of homophily and the stronger set of assumptions, the identification arguments in 4.1-4.3 remain valid, based on which we then establish the identification of the parametric homophily parameter $\beta_0$.

Under the nonparametric homophily setting and Assumptions 1, 2, 3 and 4, Subsection

4.1 first establishes the identification of pairs of individuals $(ij)$ with the same observable characteristics $(x_i = x_j = \overline{x})$ and the same unknown fixed effects $(A_i = A_j = a)$, based on the observable conditional popularity of $i$ and $j$. Subsection 4.2 then shows that we may identify the common yet unknown fixed effect of such pairs, if their common fixed effect corresponds to one of the two normalized quantiles of $F$, the unknown CDF of the idiosyncratic pairwise shocks. This identifies the unknown fixed effect $A_i = A_j = a$ for two particular values of $a$, illustrating the usefulness of the two-quantile normalization adopted in 3. Subsection 4.3 then demonstrates how to use an "in-fill" and "out-expansion" algorithm to extend the identification results for these two initial values of fixed effects to all other possible values of fixed effects. In the meanwhile, Subsection 4.3 also shows that the whole unknown CDF $F$ can also be identified along with the fixed effects $A$. Subsection 4.4 establishes the identification of $w$, the nonparametric homophily effect function.

Until Subsection 4.5, we maintain the following model specification, assumptions and normalization:

$$
\begin{gathered}
D_{ij} = \mathbf{1}\left\{ w\left(X_i, X_j\right) + A_i + A_j \geq U_{ij} \right\}, \\
Q_\alpha\left[U_{ij}\right] = 0, \quad Q_{1-\alpha}\left[U_{ij}\right] = 1, \text{ for some fixed } \alpha \in \left(0, \frac{1}{2}\right), 
\end{gathered} \tag{10}
$$

along with Assumptions 1, 2, 3 and 4. We discuss how to adapt the arguments for the parametric homophily setting in 4.5.

## 4.1 Identification of Pairs with Same Observable Characteristics and Same Fixed Effects

We now start with the identification of pairs of individuals with the same observable characteristics and the same fixed effects.

First, consider two individuals $\overline{i}, \overline{j}$ such that $x_{\overline{i}} = x_{\overline{j}} = \overline{x} \in Supp\left(X_i\right)$. This is valid for identification analysis, supposing that we can observe the whole population of the observable data. We now also provide some heuristics for the validity of this choice if a (countable) size of the sample $N$. This is clearly possible for sufficiently large $N$ if $X_i$ is discretely distributed at $\overline{x}$. If on the other hand, $X_i$ is continuously distributed at $\overline{x}$, which lies inside $Supp\left(X_i\right)$, then for $N$ large enough, we will have as many individuals $j$ as we want with characteristics $x_j$ as close to $\overline{x}$ as possible in Euclidean norm. As $w\left(\cdot,\cdot\right)$ is taken to be continuous with respect to Euclidean norm, we could use a limit argument to make this precise. However, in this paper, we focus on the identification argument as if we have perfect knowledge of the joint distribution of the observable data.

Given that we have chosen $\bar{i}, \bar{j}$ with observable and known characteristics $x_{\bar{i}} = x_{\bar{j}} = \bar{x}$ and unknown fixed effects $A_{\bar{i}}, A_{\bar{j}}$, we know that

$$w\left(x_{\bar{i}}, x_h\right) = w\left(x_{\bar{j}}, x_h\right) \quad \forall h \in N \backslash \left\{\bar{i}, \bar{j}\right\},$$

Then the conditional "popularity" of $\bar{i}$

$$\mathbb{E}\left[D_{\bar{i}h} \middle| x_{\bar{i}} = \bar{x}, A_{\bar{i}}\right] = \mathbb{P}\left\{w\left(\bar{x}, x_h\right)' \beta_0 + A_{\bar{i}} + A_h \geq U_{ih} \middle| A_{\bar{i}}\right\}$$

is strictly increasing in $A_{\bar{i}}$ by the strict monotonicity of $F$ on $\mathbb{R}$. In other words, if

$$\mathbb{E}\left[D_{\bar{i}h} \middle| x_{\bar{i}} = \bar{x}, A_{\bar{i}}, h \neq \bar{i}, \bar{j}\right] \geq \mathbb{E}\left[D_{\bar{j}h} \middle| x_{\bar{j}} = \bar{x}, A_{\bar{j}}, h \neq \bar{i}, \bar{j}\right],$$

we can infer that $A_i \geq A_j$.

We now argue that the conditional popularity $\mathbb{E}\left[D_{\bar{i}k} \middle| x_{\bar{i}} = \bar{x}, A_{\bar{i}}\right]$ is "identified from data": conditional on a fixed individual $\bar{i}$, who has fixed observable characteristics $\bar{x}$ and some fixed unobservable characteristics $A_{\bar{i}} = a$ for some unknown $a$, the indicator variable for the presence of a link between individual $\bar{i}$ and some other individual $k$ is given by

$$D_{\bar{i}k}\big|_{x_{\bar{i}} = \bar{x}, A_{\bar{i}} = a} = \mathbf{1}\left\{w\left(\bar{x}, x_k\right)' \beta_0 + a + A_k \geq U_{ik}\right\}.$$

By Assumption 2 (IID idiosyncratic pairwise shock $U_{ij}$) and Assumption 3 (random sampling of $X_k, A_k$) , the collection of random variables $\{D_{\bar{i}k}\}_{k \in N \backslash \{i\}}$ are i.i.d. conditional on the identity (i.e., the characteristics) of individual $\bar{i}$. As for identification analysis we assume that the joint distribution of observable data is known, it follows that we know the conditional distribution of observable $D_{\bar{i}k}$, conditional on the observable identity of individual $\bar{i}$.[8] Hence, the conditional popularity of individual $\bar{i}$

$$\mathbb{E}\left[D_{\bar{i}k} \middle| x_{\bar{i}} = \bar{x}, A_{\bar{i}}\right] = \mathbb{P}\left\{w\left(\bar{x}, x_h\right)' \beta_0 + A_{\bar{i}} + A_h \geq U_{ih} \middle| A_{\bar{i}}\right\}$$

is directly identified from data.

Hence, by comparing the identified popularity of any two fixed individuals $ij$ with the same observable characteristics $x_i = x_j = \bar{x}$, we have the following identifying relationships:

$$\mathbf{1}\left\{A_i \geq A_j \text{ and } x_i = x_j = \bar{x}\right\} = \mathbf{1}\left\{\mathbb{E}\left[D_{\bar{i}h} - D_{\bar{j}h} \middle| x_{\bar{i}} = x_j = \bar{x}, A_{\bar{i}}, A_{\bar{j}}, h \neq \bar{i}, \bar{j}\right] \geq 0\right\} \quad (11)$$

[8] Clearly, given that $\{D_{\bar{i}k}\}_{k \in N \backslash \{i\}}$ are conditionally i.i.d., by the Strong Law of Large Numbers, or more fundamentally the Ergodic Theorem, $\mathbb{E}\left[D_{\bar{i}k} \middle| x_{\bar{i}} = \bar{x}, A_{\bar{i}}\right]$ is consistently estimatable by the corresponding sample averages.

$$\mathbf{1}\left\{A_i = A_j \text{ and } x_i = x_j = \overline{x}\right\} = \mathbf{1}\left\{\mathbb{E}\left[D_{\bar{i}h} - D_{\bar{j}h}\middle| x_{\bar{i}} = x_j = \overline{x}, A_{\bar{i}}, A_{\bar{j}}, h \neq \bar{i}, \bar{j}\right] = 0\right\} \tag{12}$$

where the right hand sides have been argued to be identified from data. In summary, we are able to identify an ordering on the fixed effects of all individuals with the same observable characteristics $\overline{x}$ via (11).

## 4.2 Identification of Pairs with Same Observable Characteristics and Certain Fixed Effects

Now that we can identify pairs of individuals $\bar{i}, \bar{j}$ with the same observable characteristics $x_{\bar{i}} = x_{\bar{j}} = \overline{x}$ and the same fixed effects, we now fix any unknown constant $a$ and consider pairs of individuals with observable $\overline{x}$ and the same fixed effects $A_{\bar{i}} = A_{\bar{j}} = a$ for this particular $a$.

We first try to identify whether $a = 0$. This is conceptually possible by Assumption 4 (conditional full support of $A_i$): regardless of $\overline{x}$, 0 lies in the conditional support of $A_i|_{X_i=\overline{x}}$. We proceed by considering the observable frequency of links $D_{\bar{i}\bar{j}}$ between all such pairs of $\bar{i}, \bar{j}$:

$$\begin{aligned}\mathbb{E}\left[D_{\bar{i}\bar{j}}\middle| x_{\bar{i}} = x_{\bar{j}} = \overline{x},\ A_{\bar{i}} = A_{\bar{j}} = a\right] &= \mathbb{P}\left\{0 + 2a \geq U_{ij}\middle| x_{\bar{i}} = x_{\bar{j}} = \overline{x},\ A_{\bar{i}} = A_{\bar{j}} = a\right\} \\ &= F\left(2a\right) = \alpha \text{ if and only if } a = 0,\end{aligned}$$

by the normalization $Q_\alpha\left[U_{ij}\right] = 0$ and the strict monotonicity of $F$. In summary, we have identified

$$\mathbf{1}\left\{A_{\bar{i}} = A_{\bar{j}} = 0 \text{ and } x_{\bar{i}} = x_{\bar{j}} = \overline{x}\right\} = \mathbf{1}\left\{\mathbb{E}\left[D_{\bar{i}\bar{j}}\middle| x_{\bar{i}} = x_{\bar{j}} = \overline{x},\ A_{\bar{i}} = A_{\bar{j}}\right] = \alpha\right\}$$

Similarly, we then may identify whether $a = \frac{1}{2}$. We may achieve this again by considering the observable frequency of links $D_{\bar{i}\bar{j}}$ between such pairs of $\bar{i}, \bar{j}$:

$$\begin{aligned}\mathbb{E}\left[D_{\bar{i}\bar{j}}\middle| x_{\bar{i}} = x_{\bar{j}} = \overline{x},\ A_{\bar{i}} = A_{\bar{j}}\right] &= \mathbb{P}\left\{0 + 2a \geq U_{ij}\middle| x_{\bar{i}} = x_{\bar{j}} = \overline{x},\ A_{\bar{i}} = A_{\bar{j}}\right\} \\ &= F\left(2a\right) = 1 - \alpha \text{ if and only if } a = \frac{1}{2},\end{aligned}$$

so we have identified

$$\mathbf{1}\left\{A_{\bar{i}} = A_{\bar{j}} = \frac{1}{2} \text{ and } x_i = x_j = \overline{x}\right\} = \mathbf{1}\left\{\mathbb{E}\left[D_{\bar{i}\bar{j}}\middle| x_{\bar{i}} = x_{\bar{j}} = \overline{x},\ A_{\bar{i}} = A_{\bar{j}}\right] = 1 - \alpha\right\}.$$

In summary, by considering the conditional linking probability among individuals with the same observable characteristics and the same fixed effects, we are able to identify two levels of their common fixed effects, corresponding to the two normalized quantiles of the CDF $F$.

## 4.3 Identification of the CDF $F$ and the Fixed Effects $A$

We now proceed to establish the identification of $F$ on its whole domain $\mathbb{R}$ and $A_i$ for every possible value in $\mathbb{R}$.

So far, either by normalization or by identification, we have pinned down the values of $F$ and $A_i$ on

$$\mathcal{F}_0 = \{0, 1\}, \quad \mathcal{A}_0 = \left\{0, \frac{1}{2}\right\},$$

i.e., we know the values of $F(0), F(1)$, $\mathbf{1}\{A_i = 0\}$, and $\mathbf{1}\left\{A_i = \frac{1}{2}\right\}$.

**"In-Fill" Algorithm**

Using this knowledge, we first carry out an "in-fill" algorithm that identifies $F$ and $A$ on all bisection points in $[0, 1]$. To start with, we still focus on individuals with the same observable characteristics $\overline{x}$, but now we can select pairs that have two different known levels of fixed effects, "0" and "$\frac{1}{2}$", respectively, and then consider the conditional linking probability across these two levels of fixed effects, leading to the identification of $F\left(\frac{1}{2}\right)$ via

$$F\left(\frac{1}{2}\right) = \mathbb{P}\left\{w(\overline{x}, \overline{x}) + 0 + \frac{1}{2} \geq U_{ij}\right\} = \mathbb{E}\left[D_{ij} \middle| x_i = x_j = \overline{x},\ A_i = 0, A_j = \frac{1}{2}\right]$$

Then, we may extend the results to back out more unknown levels of fixed effects:

$$\begin{aligned}
\mathbf{1}\left\{A_i = A_j = \frac{1}{4}\right\} &= \mathbf{1}\left\{\mathbb{E}\left[D_{ij} \middle| x_i = x_j = \overline{x},\ A_i = A_j\right] = F\left(\frac{1}{2}\right)\right\}, \\
\mathbf{1}\left\{A_j = \frac{3}{4}\right\} &= \mathbf{1}\left\{\mathbb{E}\left[D_{ij} \middle| x_i = x_j = \overline{x},\ A_i = \frac{1}{4}, A_j\right] = F(1)\right\}, \\
\mathbf{1}\{A_j = 1\} &= \mathbf{1}\{\mathbb{E}[D_{ij} | x_i = x_j = \overline{x},\ A_i = 0, A_j] = F(1)\}.
\end{aligned}$$

The first equation reiterates the identification argument in the Subsection 4.1 based on conditional linking probabilities among individuals with same fixed effects. The latter two equations utilize a triangulation method, where we control one known level of fixed effect $A_i$ and a known value of the CDF $F$, and then back out another level of unknown fixed effect $A_j$ by considering conditional linking probabilities among individuals with different fixed effects. In summary of the four identification equations above, we now have identified

$F$ and $A_i$ on

$$\mathcal{F}_1 = \left\{0, \frac{1}{2}, 1\right\}, \quad \mathcal{A}_1 = \left\{0, \frac{1}{4}, \frac{1}{2}, \frac{3}{4}, 1\right\},$$

Inductively, suppose that $F$ and $A_i$ have been identified on

$$\mathcal{F}_n := \left\{\frac{m}{2^n} : m \leq 2^n\right\}, \quad \mathcal{A}_n := \left\{\frac{m}{2^{n+1}} : m \leq 2^{n+1}\right\}.$$

Then we may apply the same reasoning and further identify

$$F\left(\frac{m+h}{2^{n+1}}\right) = \mathbb{E}\left[D_{ij} \middle| x_i = x_j = \overline{x},\ A_i = \frac{m}{2^{n+1}}, A_j = \frac{h}{2^{n+1}}\right] \quad \forall m, h \leq 2^n,$$

$$\mathbf{1}\left\{A_i = A_j = \frac{m}{2^{n+2}}\right\} = \mathbf{1}\left\{\mathbb{E}\left[D_{ij} \middle| x_i = x_j = \overline{x},\ A_i = A_j\right] = F\left(\frac{m}{2^{n+1}}\right)\right\} \quad \forall m \leq 2^{m+1},$$

$$\mathbf{1}\left\{A_j = 1 - \frac{m}{2^{n+2}}\right\} = \mathbf{1}\left\{\mathbb{E}\left[D_{ij} \middle| x_i = x_j = \overline{x},\ A_i = \frac{m}{2^{n+2}}, A_j\right] = F(1)\right\} \quad \forall m \leq 2^{m+1}.$$

establishing further identification of $F$ and $A_i$ on $\mathcal{F}_{n+1}$ and $\mathcal{A}_{n+1}$. By induction, we can thus identify $F$ and $A$ on all "bisection points" of $[0,1]$:

$$\mathcal{F}_\infty = \left\{\frac{m}{2^n} : m \leq 2^n, n \in \mathbb{N}\right\}, \quad \mathcal{A}_\infty = \left\{\frac{m}{2^n} : m \leq 2^n, n \in \mathbb{N}\right\}.$$

We now extend the identification results on all bisection points of $[0,1]$ to the whole real interval $[0,1]$ by taking limits. Consider any real number $a \in [0,1] \backslash \mathcal{A}_\infty$. Then there exist two sequences $(\underline{a}_n), (\overline{a}_n)$ in $\mathcal{A}_\infty$ such that $\underline{a}_n \nearrow a$ and $\overline{a}_n \searrow a$ as $n \to \infty$. Recall that we can identify an order "$\geq$" on "fixed-effect equivalent class with observable $\overline{x}$". For each $n \in \mathbb{N}$, inequalities of the form $\mathbf{1}\{A_i \in [\underline{a}_n, \overline{a}_n]\}$ are identified. Hence,

$$\mathbf{1}\{A_i = a\} = \prod_{n=1}^{\infty} \mathbf{1}\{A_i \in [\underline{a}_n, \overline{a}_n]\} = \mathbf{1}\left\{A_i \in \bigcap_{n=1}^{\infty} [\underline{a}_n, \overline{a}_n]\right\}$$

is identified. To identify $F(a)$ for $a \in [0,1] \backslash \mathcal{A}_\infty$, we can simply take the upper limit[9]

$$F(a) = \lim_{\overline{a}_n \searrow a} F(\overline{a}_n).$$

We have thus established identification of $F$ and $A$ on $\overline{\mathcal{F}} = [0,1]$ and $\overline{\mathcal{A}} = [0,1]$, with the range of values $F\left(\overline{\mathcal{F}}_0\right) = [F(0), F(1)] = [\alpha, 1-\alpha]$ as imposed by the two quantile normalization.

[9] Recall that a CDF is always right-continuous, so the upper limit $F(a) = \lim_{\overline{a}_n \searrow a} F(\overline{a}_n)$ also holds even if $F$ is not continuous. Moreover, if we allow $F$ to be discontinuous, we can also identify the lower limit by $F_-(a) = \lim_{\underline{a}_n \nearrow a} F(\overline{a}_n)$.

**"Out-Expansion" Algorithm**

The "in-fill" algorithm above establishes identification on the unit interval $[0,1]$, we now switch to an "out-expansion" algorithm that further extends the identification results from $[0,1]$ to $(-\infty,\infty)$.

Using a similar reasoning as above, we have

$$\begin{aligned}\mathbf{1}\{A_j=-a\}=\mathbf{1}\{\mathbb{E}[D_{ij}|\,x_i=x_j=\overline{x},\ A_i=a,A_j]=\alpha\}\quad&\forall a\in[0,1]\\ F(-a)=\mathbb{E}[D_{ij}|\,x_i=x_j=\overline{x},\ A_i=0,A_j=-a]\quad&\forall a\in[0,1]\end{aligned}$$

and thus we have identified $A_i$ and $F$ on $\overline{\mathcal{A}}_0=[-1,1]$ and $\overline{\mathcal{F}}_0=[-1,1]$.

Inductively, suppose that $A_i$ and $F$ have already been defined on $\overline{\mathcal{A}}_n=\overline{\mathcal{F}}_n\supseteq[-2^n,2^n]$ for any $n\in\mathbb{N}$. Then we can further identify

$$\begin{aligned}F(a+b)&=\mathbb{E}[D_{ij}|\,x_i=x_j=\overline{x},\ A_i=a,A_j=b],\ \forall a,b\in[-2^n,2^n]\\ \mathbf{1}\{A_j=a\}&=\mathbf{1}\{\mathbb{E}[D_{ij}|\,x_i=x_j=\overline{x},\ A_i=b,A_j=a]=c\}\\ &\qquad\qquad\forall b\in[-2^n,2^n],\forall c\in[F(-2^n),F(2^n)]\end{aligned}$$

Thus we have further identify $F$ and $A_i$ on

$$\begin{aligned}\overline{\mathcal{F}}_{n+1}&=\{a+b:a,b\in[-2^n,2^n]\}=\left[-2^{n+1},2^{n+1}\right]\\ \overline{\mathcal{A}}_{n+1}&=\left\{F^{-1}(c)-b:\forall b\in[-2^n,2^n],\forall c\in[F(-2^n),F(2^n)]\right\}\\ &=[-2^n,2^n]-[-2^n,2^n]=\left[-2^{n+1},2^{n+1}\right]\end{aligned}$$

By induction, we can identify $F$ and $A_i$ on $\bigcup_{n=1}^{\infty}\overline{\mathcal{F}}_n=\bigcup_{n=1}^{\infty}\overline{\mathcal{A}}_n=\mathbb{R}$.

## 4.4 Identification of the Homophily Effect Function $w$

We now continue to identify $w$ on its domain $Supp(X_i)\times Supp(X_i)$. Recall that $A_i$ and $F$ have all been identified in Subsection 4.3, so we may treat them as known.

We consider pairs of individuals with any observable characteristics $(x_i,x_j)\in Supp(X_i,X_j)$ and zero fixed effects $A_i=A_j=0$. By

$$F(w(x_i,x_j))=\mathbb{P}\{w(x_i,x_j)\geq U_{ij}\}=\mathbb{E}[D_{ij}|\,x_i\neq x_j,\ A_i=A_j=0]$$

we can identify

$$w(x_i,x_j)=F^{-1}(\mathbb{E}[D_{ij}|\,x_i\neq x_j,\ A_i=A_j=0]).$$

Hence, we have identified $w,F,A$ up to the normalization we made. Theorem 1 translates the point identification under normalization proved here back into a set identification result, where the identified set is characterized by a specific form of positive affine transformations. We thus have proved Theorem 1 under Assumptions 1, 2, 3 and 4.

### 4.5 Identification of the Homophily Effect Parameter $\beta_0$

We now switch to the parametric homophily setting, and consider the identification of the linear homophily effect parameter $\beta_0$ in (4) under Assumptions 1a, 2, 3, 4 and 5a. Despite the fact that identification under the nonparametric homophily setting does not imply identification under the parametric homophily setting, the identification arguments in previous subsections are very useful in the parametric homophily setting.

We maintain the normalization $Q_\alpha [U_{ij}] = 0,\ Q_{1-\alpha} [U_{ij}] = 1$. By Assumption 1a, $\theta = w(\overline{x}, \overline{x}) = \tilde{w}(\overline{x}, \overline{x})' \beta_0 = 0$, which is consistent with the previous location normalization $\theta = 0$ in the identification analysis under the nonparametric homophily setting. Moreover, as Assumption 1a, 2, 3, 4 and 5a imply Assumption 1, 2, 3 and 4 by Lemma 2, all previous identification results carry over without change.

By the result in Subsection 4.3, $A_i$ and $F$ are identified on $\mathbb{R}$. By Assumption 5a, $Supp\left(\tilde{W}_{ij}\right) = Supp\left(\tilde{w}(X_i, X_j)\right)$ contains $k$ linearly independent nonzero vectors $\overline{W}_m \in \mathbb{R}^k$ for $m = 1, ..., k$. Hence, by

$$F\left(\overline{W}_m' \beta_0\right) = \mathbb{P}\left\{\overline{W}_m' \beta_0 \geq U_{ij}\right\} = \mathbb{E}\left[D_{ij} \middle| x_i \neq x_j,\ A_i = A_j = 0\right],$$

we can identify

$$\overline{W}_m' \beta_0 = \xi_m := F^{-1}\left(\mathbb{E}\left[D_{ij} \middle| x_i \neq x_j,\ A_i = A_j = 0\right]\right).$$

Writing $\overline{W} = \left(\overline{W}_m'\right)_{m=1}^k$ and $\xi = (\xi_m)_{m=1}^k$, we have $\overline{W}' \beta_0 = \xi$. As $\overline{W}$ is invertible by construction, we can identify

$$\beta_0 = \overline{W}^{-1} \xi,$$

the right hand side of which are all known.

We have thus completed the proof of Theorem 2, establishing identification for the parametric homophily setting under Assumptions 1a, 2, 3, 4 and 5a.

# 5 Extensions

## 5.1 Nonlinear Coupling of Fixed Effects

In this section, we revisit model (9) in Remark 2:

$$D_{ij} = \mathbf{1}\left\{w\left(X_i, X_j\right) + \phi\left(A_i, A_j\right) \geq U_{ij}\right\}.$$

where $w$ is nonparametrically specified and $\phi : \mathbb{R}^2 \to \mathbb{R}$ is a *known* potentially nonlinear function.

**Assumption. 6** (Nice Coupling Function) $\phi : \mathbb{R}^2 \to \mathbb{R}$ *is a symmetric, differentiable, strictly increasing (in both arguments) and surjective (in both arguments) function.*

Assumption 6 mainly ensures that the coupled fixed effects are still "smoothly ordered" with a "full support": continuity rules out jumps in the coupled effects of $A_i$ and $A_j$ on linking probabilities, strict monotonicity preserves the implications of individual fixed effects on relative popularity, and surjectivity implies that given an arbitrary $A_i$, the other individual's fixed effect $A_j$ is still pivotal in the determination of the coupled effect.

As discussed in Remark 2, setting the two quantiles $Q_\alpha, Q_{1-\alpha}$ to $0, 1$ are no longer valid normalization. Nevertheless, we may write

$$Q_\alpha\left[U_{ij}\right] = q_\alpha, \ Q_{1-\alpha}\left[U_{ij}\right] = q_{1-\alpha}$$

for $\alpha \in \left(0, \frac{1}{2}\right)$ and two *unknown* constants $q_\alpha < q_{1-\alpha}$, and treat $q_\alpha, q_{1-\alpha}$ effectively as location and scale parameters that we seek to identify.

We show in Appendix C how to formally establish the identification results under this setting, and provide two general levels of conditions related to the identifiability (and normalizability) of the location and the scale parameters. We summarize the main findings here.

In Subsection C.1, we provide some general discussion on how the identification arguments in Subsections 4.1-4.3 can be adapted accordingly. The key difference after the adaption is that the two initially "identified" levels of fixed effects, instead of being labeled as "0" and "1" under the two-quantile normalization imposed in Section 3, must now be written as

$$\overline{\phi}^{-1}\left(q_\alpha\right), \overline{\phi}^{-1}\left(q_{1-\alpha}\right),$$

where $\overline{\phi} : \mathbb{R} \to \mathbb{R}$, defined by $\overline{\phi}\left(a\right) := \phi\left(a, a\right)$ for all $a \in \mathbb{R}$, is a *known*, potentially nonlinear, symmetric, continuous, strictly increasing and surjective function given Assumption 6, thus admitting an inverse function $\overline{\phi}^{-1}$. Even though the two numbers $\overline{\phi}^{-1}(q_\alpha), \overline{\phi}^{-1}(q_{1-\alpha})$ are unknown, we can nevertheless identify pairs of individuals who share the same unknown fixed effects $\overline{\phi}^{-1}(q_\alpha)$ or $\overline{\phi}^{-1}(q_{1-\alpha})$. Given these two initial levels of identified fixed effects, the "in-fill and out-expansion" algorithms again apply, establishing the joint identification of the individual fixed effects $A_i$ and the CDF $F$ on the whole real line. However, the remaining subtlety lies in that the identification results are all described relative to the two unknown numbers $q_\alpha, q_{1-\alpha}$, which effectively "encrypts" the whole real line. Whether the "decryption keys" $q_\alpha, q_{1-\alpha}$ can be identified depend on the curvature of $\phi$.

To illustrate this, in Subsection C.2 we solve out a concrete example, in which $\phi$ is assumed to take a cubic form:

$$\phi(a_1, a_2) := (a_1 + a_2)^3, \quad \forall a_1, a_2 \in \mathbb{R}.$$

The homogeneity of $\phi$ renders it impossible to identify the scale parameter, but in the meanwhile we may consequently impose the interquantile-range scale normalization . With one location parameter that is unknown and not normalizable, we show how to exploit the fact that $\overline{\phi}'(a) = 24a^2 = 0$ if and only if $a = 0$ to recover the location parameter via a "conjecture and falsification" argument, which illustrates the main technical method for this section and is more widely applicable beyond this example. Moreover, we discuss why Assumption 1 (degeneration of homophily effect on the hyper-diagonal) is no longer required in this case, as the identified location parameter actually endows "0" with an invariant property across different values of observable characteristics $\overline{x} \in Supp(X_i)$, so that we may obtain identification results across different $\overline{x} \in Supp(X_i)$ without Assumption 1. The precise characterization of the identification result obtained for this illustrating example is given by the following proposition.

**Proposition 1** (Nonparametric Identification with Cubic Coupling of Fixed Effects)**.** *Consider the nonparametric model (9) with cubic coupling of fixed effects: $\phi(a_1, a_2) := (a_1 + a_2)^3$ for all $a_1, a_2 \in \mathbb{R}$. Under Assumption 2, 3, 4 and 6, along with a further assumption that $F$ is differentiable at* 0*, the following set*

$$\left\{ \left( cw + b,\ c^{\frac{1}{3}} A,\ cF^{-1} + b \right) : c \in (0, \infty), b \in \mathbb{R} \right\}$$

*where $(w, A, F^{-1})$ denote the true values is identified, and no proper nonempty subset of it can be further identified.*

Proposition 1 provides a sharper characterization of the identified set than Theorem 1, as the identified set in Proposition 1 features two degrees of freedom instead of three as in Theorem 1. This is not surprising given the well-known observation in the econometric literature that nonlinearity may help with identification. Here, the curvature of $\phi$ may provide information about the location parameter, which is not identifiable under linear aggregation of fixed effects. Hence, Theorems 1 and 2, along with their proofs in previous sections, may be regarded as results for the "worst-case scenario".

Subsections C.3 and C.4 then provide two generalized forms of conditions related to homogeneity or translatability of the coupling function $\phi$, under which a similar argument as in the cubic case may be applied.

**Definition 1** (Homogeneity and Translatability ).

- $\phi$ is *generalized homogeneous* if, for any $c > 0$, there exists some (known) bijective function $g_c$ such that $\phi(g_c(a_1), g_c(a_2)) = c\phi(a_1, a_2)$, $\forall a_1, a_2 \in \mathbb{R}$.
- $\phi$ is *generalized translatable* if, for any $b \in \mathbb{R}$, there exists a bijection $h_b : \mathbb{R} \to \mathbb{R}$ such that $\phi(h_b(a_1), h_b(a_2)) = \phi(a_1, a_2) + b$, $\forall a_1, a_2 \in \mathbb{R}$.
- $\phi$ is *periodic homogeneous with period* $T > 1$ if there exists a bijective function $g_T$ such that $\phi(g_T(a_1), g_T(a_2)) = T\phi(a_1, a_2)$, $\forall a_1, a_2 \in \mathbb{R}$.
- $\phi$ is *periodic translatable with period* $T > 0$ if there exists a bijective function $h_T : \mathbb{R} \to \mathbb{R}$ such that $\phi(h_T(a_1), h_T(a_2)) = \phi(a_1, a_2) + T$, $\forall a_1, a_2 \in \mathbb{R}$.

Subsections C.3 and C.4 provide concrete examples of functions that satisfy the conditions defined above, and discuss how these conditions are sufficient for the *unidentifiability* of scale and location parameters.

**Lemma 1 (Identification with Potentially Nonlinear Coupling of Fixed Effects).**

*(i) If $\phi$ is generalized or periodic homogeneous, then the scale of the distribution of $U_{ij}$ (for instance, the interquantile range $Q_{1-\alpha}[U_{ij}] - Q_\alpha[U_{ij}]$) is unidentified.*

*(ii) If $\phi$ is generalized or periodic translatable, then the location of the distribution of $U_{ij}$ (for instance, its $\alpha$-th quantile $Q_\alpha[U_{ij}]$), is unidentified.*

Note that, in either of the cases in Lemma 1, generalized homogeneity or generalized translatability helps characterize the identified set via the corresponding induced modeling equivalence classes, which in turn helps characterize the identified invariant properties within each modeling equivalence class. If both generalized homogeneity and generalized translatability are satisfied, then the characterization of the identified sets as in Theorem 1 essentially applies (with some minor corresponding adaptions).

## 5.2 Unknown Coupling of Fixed Effects

As a further extension, we now consider a model with unknown coupling of fixed effects:

$$D_{ij} = \mathbf{1}\left\{w\left(X_i, X_j\right) + \phi\left(A_i, A_j\right) \geq U_{ij}\right\}$$

where $\phi : \mathbb{R}^2 \to \mathbb{R}$ is an *unknown* function that satisfies Assumption 6, i.e., $\phi$ is continuous, differentiable, strictly increasing and surjective in each of its arguments.

Then, take any strictly positive scalar $a \in \mathbb{R}_{++}$, any scalars $b, c \in \mathbb{R}$ and any continuous, differentiable, strictly increasing and surjective function $f : \mathbb{R} \to \mathbb{R}$, and define

$$\begin{aligned} \hat{A}_i &:= f\left(A_i, \right), \\ \hat{\phi}\left(\hat{A}_i, \hat{A}_j\right) &:= c\phi\left(f^{-1}\left(\hat{A}_i\right), f^{-1}\left(\hat{A}_j\right)\right) + a \\ \hat{w}\left(x_i, x_j\right) &:= cw\left(x_i, x_j\right) + b, \\ \hat{U}_{ij} &:= cU_{ij} + a + b. \end{aligned}$$

The above transformation of the unknown $(w, \phi, A, U)$ again maintains modeling equivalence. In particular, $\hat{\phi}$ remains symmetric, continuous, differentiable, strictly increasing and surjective, so that Assumption 6 is preserved. Hence, we may without loss of generality impose the following normalization:

$$\theta \equiv w\left(x_i, x_i\right) = 0, \quad Q_\alpha\left[U_{ij}\right] = 0, \quad Q_{1-\alpha}\left[U_{ij}\right] = 1, \text{ for some } \alpha \in \left(0, \tfrac{1}{2}\right)$$
$$\overline{\phi}\left(a\right) := \phi\left(a, a\right) = 2a \text{ for all } a \in \mathbb{R}.$$

To obtain the last normalization on $\overline{\phi}$ in particular, we may take $f\left(a\right) := \frac{1}{2}\overline{\phi}\left(a\right) := \frac{1}{2}\phi\left(a, a\right)$ in the transformation defined above, which ensures that the transformed coupling function satisfies $\hat{\phi}\left(a, a\right) = \overline{\phi}\left(f^{-1}\left(a\right)\right) = \overline{\phi}\left(\overline{\phi}^{-1}\left(2a\right)\right) = 2a$.

Notice that all the normalization imposed in Section 3 are still imposed here. Moreover, when individuals share the same fixed effects, the last normalization on $\phi$ takes us essentially back to the case with linearly additive fixed effects when the fixed effects of two individuals coincide. A similar argument as provided in Subsection 5.1 (and Appendix C) can again be applied here, but now all unknown and unnormalized values of $\phi\left(a_1, a_2\right)$ for $a_1 \neq a_2$ remain to identified to "decrypt" the identification results obtained via the "in-fill and out-expansion" algorithm. Complicated as it may seem, this problem is still similar to the one encountered in the previous subsection where $\phi$ is known but nonlinear, which may again be approached via the "conjecture and falsification" argument illustrated in the example of cubic coupling function. See Appendix D for a more detailed discussion.

### 5.3 Nonseparability of Homophily and Degree Heterogeneity

We now consider a further generalization where nonparametric dependence between the homophily effects and the fixed effects is allowed. Formally, we consider the following model

$$D_{ij} = \mathbf{1}\left\{\phi\left(X_i, X_j; A_i, A_j\right) \geq U_{ij}\right\}, \tag{13}$$

where $\phi : \mathbb{R}^k \times \mathbb{R}^k \times \mathbb{R} \times \mathbb{R} \to \mathbb{R}$ is an unknown function. This setting generalizes the link formation model considered by Johnsson and Moon (2017) of the form:

$$D_{ij} = \mathbf{1}\left\{\phi\left(w\left(X_i, X_j\right); A_i, A_j\right) \geq U_{ij}\right\}. \tag{14}$$

**Assumption. 6'**(Nice Index Function) $\phi\left(x_1, x_2; a_1, a_2\right)$ *is a symmetric with respect to* $\left(x_1, a_1\right)$ *and* $\left(x_2, a_2\right)$[10]*, differentiable, and strictly increasing in* $a_1$ *and* $a_2$*.*

For technical clarity, we replace surjectivity of $\phi$ and conditional full support of $A_i$ with the following joint assumption on $\phi$ and $A_i$:

**Assumption. 4'**(Conditional Pivotality of $A_i$) $\forall x_1, x_2, a_1,$ we have $Supp\left(A_i \middle| X_2 = x_2\right) \equiv Supp\left(A_i\right)$ and $F\left(\phi\left(x_1, x_2; a_1, Supp\left(A_i\right)\right)\right) = (0, 1)$.

We impose the following key assumption, which is an adaption of Assumption 1:

**Assumption. 1"** (Degeneration of Homophily Effect on the Joint-Hyper-Diagonal) $\forall \overline{x}, \overline{x}' \in Supp\left(X_i\right), \forall a \in Supp\left(A_i\right),\ \phi\left(\overline{x}, \overline{x}; a, a\right) = \phi\left(\overline{x}', \overline{x}'; a, a\right) =: \overline{\phi}_A\left(a\right).$

Essentially, Assumption 1" basically says that, the linking probability among individuals who share the same observable characteristics $x_i = x_j = \overline{x}$ and the same fixed effects $a_1 = a_2 = a$ do not depend on the level of $\overline{x}$, which embeds the economic interpretation of homophily effects. Assumption 1" is more likely to be reasonable if we are mostly interested in "distance-based" characteristics. In particular, Assumption 1" is implied by Assumption 1 under model (14) as considered in Johnsson and Moon (2017), where observable characteristics are first coupled into an homophily effect index. Besides, Assumption 1" is clearly implied by Assumption 1 under the additively separable model (30) considered in previous sections. On another aspect, Assumption 1" is also weaker than Assumption 1 in the sense that it does not require that $\phi\left(\overline{x}, \overline{x}; a_1, a_2\right)$ be independent of $\overline{x}$ when $a_1 \neq a_2$.

Given Assumption 1", 2, 3, 4' and 6', we now present the final result:

**Theorem 3** (Nonparametric Identification under Nonseparability)**.** *For the nonseparable index setting (13), under Assumptions 1", 2, 3, 4' and 6',* $\phi$ *is identified up to normalization.*

[10]Precisely, $\phi\left(x_1, x_2; a_1, a_2\right) \equiv \phi\left(x_2, x_1; a_2, a_1\right).$

*Proof.* Taking any strictly increasing function $g, f$, we define the following transformation of the unknown $(U, A, \phi)$:

$$\hat{U}_{ij} := g\left(U_{ij}\right), \quad \hat{A}_i := f\left(A_i\right),$$
$$\hat{\phi}\left(x_1, x_2; \hat{a}_1, \hat{a}_2\right) := g\left(\phi\left(x_1, x_2; f^{-1}\left(\hat{a}_1\right), f^{-1}\left(\hat{a}_2\right)\right)\right)$$

It is straightforward to check that modeling equivalence is maintained. In particular, as

$$\begin{aligned}\hat{\phi}\left(X_i, X_j; \hat{A}_i, \hat{A}_j\right) &= g\left(\phi\left(X_i, X_j; f^{-1}\left(f\left(A_i\right)\right), f^{-1}\left(f\left(A_j\right)\right)\right)\right) \\ &= g\left(\phi\left(X_i, X_j; A_i, A_j\right)\right) \\ &\geq \hat{U}_{ij} = g\left(U_{ij}\right) \quad \text{iff} \quad \phi\left(X_i, X_j; A_i, A_j\right) \geq U_{ij}.\end{aligned}$$

Moreover, Assumption 6' and 4' are obviously preserved as well with the transformed CDF $\hat{F} := F \circ g^{-1}$.

Taking $g\left(u\right) := F\left(u\right),\ f\left(a\right) := F\left(\overline{\phi}_A\left(a\right)\right)$, we impose the following normalizing transformation:

$$\begin{aligned}\hat{U}_{ij} &:= F\left(U_{ij}\right) \sim U\left[0,1\right] \\ \hat{A}_i &:= F\left(\overline{\phi}_A\left(A_i\right)\right) \text{ has support } (0,1) \\ \hat{\phi}\left(x_1, x_2; \hat{a}_1, \hat{a}_2\right) &:= F\left(\phi\left(x_1, x_2; \overline{\phi}_A^{-1}\left(F^{-1}\left(\hat{a}_1\right)\right), \overline{\phi}_A^{-1}\left(F^{-1}\left(\hat{a}_2\right)\right)\right)\right)\end{aligned} \tag{15}$$

As a result of the normalization, we have

$$\begin{aligned}\overline{\hat{\phi}}_A\left(\hat{a}\right) := \hat{\phi}\left(\overline{x}, \overline{x}; \hat{a}, \hat{a}\right) &= F\left(\overline{\phi}_A\left(\overline{\phi}_A^{-1}\left(F^{-1}\left(\hat{a}\right)\right)\right)\right) \\ &= F\left(F^{-1}\left(\hat{a}\right)\right) = \hat{a}.\end{aligned} \tag{16}$$

In other words, we can without loss of generality normalize the model so that the function $\phi$ reduces completely to identity on the joint-hyper-diagonal, $(x_i, a_i) = (x_j, a_j)$. Relative to the additive separable specifications considered in previous sections, where only two quantiles of $F$ may be normalized, now the whole CDF are effectively normalized, which is not surprising given the generality of the current specification and the strength of Assumption 1".

Identification of what remains unknown is then straightforward. As we no longer need to identify $F$, we may simply focus on identifying the fixed effects by considering the linking probability among individuals with the same observable characteristics $\overline{x}$ and the same

(unknown) level of fixed effects $a$. By (16), for any $a \in (0,1)$ we have

$$\mathbf{1}\{A_i = A_j = a\} = \mathbf{1}\{a = \mathbb{E}[D_{ij}|\, x_i = x_j = \overline{x}, A_i = A_j]\},$$

with the right-hand side directly identified from data.

With the fixed effects $A_i$ all identified, we select individuals with arbitrary observable characteristics $x_1, x_2$ and arbitrary known fixed effects $a_1, a_2$, and then we have

$$\phi(x_1, x_2; a_1, a_2) = \mathbb{E}[D_{ij}|\, x_i = x_1, x_j = x_2, A_i = a_1, A_j = a_2].$$

Hence, $\phi$ is identified up to the normalization we made in (15). □

Finally, it should be pointed out that the result in this section under the highly general model (13) does not imply the identification results obtained in previous sections with additively separable index. Though additively separable index models are special cases of the general nonseparable model, the stronger assumption of additive separability does allow us to obtain stronger identification results: the linear internal structure imposed by additive separability between the homophily effect index and the fixed effect index is restrictive so that we only have two degrees of freedom in terms of location and scale across the two indexes. The nonseparable model (13), with weaker modeling specification and assumptions, only allows weaker result of identification.

## 5.4 Sparse Networks

So far we have been focusing on the case of dense networks, where conditional linking probabilities of the form

$$\mathbb{E}[D_{ij}|\, X_i = x_1, X_j = x_2, A_i = a_1, A_j = a_2]$$

is guaranteed to be strictly positive by Assumption 2. The nondegenerate conditional linking probabilities then allow us to identify a complete ordering of fixed effects as well as to carry out the "in-fill and out-expansion algorithm" in Section 4.

A form of network sparsity, however, may be accommodated by augmenting the model with a stochastic meeting process, which, along with its several mathematically equivalent variants, has been used in the literature to control scale of linking probabilities in network and matching models, such as Menzel (2015), Graham (2017), Mele (2017a,b).

Specifically, consider a stochastic meeting process where any pair of individuals $ij$ ran-domly meet with a strictly positive probability $p_n$ such that

$$p_n \to 0, \quad \text{as } n \to \infty.$$

The rate at which $p_n$ vanishes asymptotically is often used as measure of network sparsity, and the identification arguments presented in previous sections is adaptable to the case where

$$np_n \to \infty, \quad \text{as } n \to \infty.$$

For simplicity, assume that $p_n = Cn^{-\kappa} + o\left(n^{-\kappa}\right)$ for some fixed $\kappa \in (0,1)$ so that

$$n^{\kappa} p_n \to C \quad \text{as } n \to \infty.$$

Let $M_{ij}$ be an indicator variable for whether $ij$ meet, with $M_{ij} = 1$ denotes the event that $ij$ meet. Conditional on the event $M_{ij} = 1$, individuals $ij$ then decide whether to form a link according to our main model (3). In summary, we could write the full network structure as

$$D_{ij} = M_{ij} \cdot \mathbf{1}\left\{w\left(X_i, X_j\right) + A_i + A_j \geq U_{ij}\right\}.$$

Supposing that the stochastic meeting process $M$ is independent of other variables $(X, A, U)$, we may then write

$$\mathbb{E}_n\left[D_{ij} \middle| X_i = x_1, X_j = x_2, A_i = a_1, A_j = a_2\right] = p_n \cdot F\left(w\left(x_1, x_2\right) + a_1 + a_2\right).$$

Note the left-hand side now should not be treated as an observable quantity that is directly identified from data, as for any finite $n$, observation of a single network is insufficient to recover the true conditional moment $\mathbb{E}_n$. Moreover, as $n \to \infty$, $p_n$ and $\mathbb{E}_n$ changes with it, so $\mathbb{E}_n$ is not a proper quantity for identification analysis.

To circumvent this problem, write

$$\begin{aligned} n^{\kappa}\mathbb{E}_n\left[D_{ij} \middle| X_i = x_1, X_j = x_2, A_i = a_1, A_j = a_2\right] &= n^{\kappa} p_n \cdot F\left(w\left(x_1, x_2\right) + a_1 + a_2\right). \\ &\to C \cdot F\left(w\left(x_1, x_2\right) + a_1 + a_2\right) \end{aligned}$$

and consider, for example as in Subsection 4.1, two specific individuals $i$ and $j$ with observable characteristics $X_i = X_j = \overline{x}$ and unknown fixed effects $A_i = a_i$, $A_j = a_j$. Then

$$\mathbb{E}_n\left[n^{\kappa} D_{ik} \middle| X_i = \overline{x}, A_i = a_i\right] \to C \cdot \mathbb{E}\left[F\left(w\left(\overline{x}, X_k\right) + a_i + A_k\right)\right]$$

and thus by the triangular-array law of large numbers we have

$$\frac{1}{n}\sum_{k\neq i} n^{\kappa} D_{ik} \xrightarrow{p} C\cdot\mathbb{E}\left[F\left(w\left(\overline{x},X_k\right)+a_i+A_k\right)\right],$$

which not only establishes the identification of $\kappa$ as the unique number such that the right-hand side is nondegenerate to 0 or 1, but also identifies the right-hand-side quantity $C\cdot\mathbb{E}\left[F\left(w\left(\overline{x},X_k\right)+a_i+A_k\right)\right]$. Then, we have

$$\mathbf{1}\left\{a_i\geq a_j\right\}=\mathbf{1}\left\{\frac{C\cdot\mathbb{E}\left[F\left(w\left(\overline{x},X_k\right)+a_i+A_k\right)\right]}{C\cdot\mathbb{E}\left[F\left(w\left(\overline{x},X_k\right)+a_j+A_k\right)\right]}\geq 1\right\}.$$

establishing the identification of the ordering of fixed effects.

Proceeding to consider pairs of individuals with $X_i=X_j=\overline{x}$ and the same fixed effects $A_i=A_j=a$ as in Subsection 4.2,

$$\frac{1}{\#\left\{(i,j):\begin{array}{cc} i\neq j, & X_i=X_j=\overline{x}, \\ 1\leq i,j\leq n & A_i=A_j=a.\end{array}\right\}}\sum_{\substack{i\neq j:\\ X_i=X_j=\overline{x},\\ A_i=A_j=a}}^{n} n^{\kappa}D_{ij}\xrightarrow{p} C\cdot F\left(2a\right).$$

giving identification of $C\cdot F\left(2a\right)$. By Assumption 4 (Conditional Full Support of $A_i$), we may identify $C$ by

$$C=\sup_{a\in\mathbb{R}}\left\{C\cdot F\left(2a\right)\right\},$$

as we have identified a complete ordering of $a$. With $C$ identified, $F\left(2a\right)$ can be identified and then inverted at the two normalized quantiles by

$$F\left(2a\right)=\begin{cases}\alpha, & \text{if and only if } a=0,\\ 1-\alpha, & \text{if and only if } a=\frac{1}{2}.\end{cases}$$

All remaining identification arguments, including the "in-fill and out-expansion" algorithm, carry over exactly as in Section 4.

More generally, the meeting probabilities may be contingent on individual characteristics, for example, in the following form:

$$p_n\left(X_i,X_j\right)=p_n\cdot q\left(X_i,X_j\right)$$

which naturally accommodates heterogeneity in meeting probabilities. Again we assume

$p_n = n^{-\kappa} + o\left(n^{-\kappa}\right)$. Here we implicitly normalize the constant before $n^{-\kappa}$ to be $C = 1$, given that we may otherwise always redefine $\tilde{q}\left(X_i, X_j\right) := Cq\left(X_i, X_j\right)$. It is then straightforward to check that again a complete ordering of $A_i$ and $A_j$ can be identified among individuals who share the same observable characteristics. Then, considering pairs of individuals with $X_i = X_j = \overline{x}$ and the same fixed effects $A_i = A_j = a$, we can identify $q\left(\overline{x}, \overline{x}\right) \cdot F\left(2a\right)$. We now can identify

$$q\left(\overline{x}, \overline{x}\right) = \sup_{a \in \mathbb{R}} \left\{q\left(\overline{x}, \overline{x}\right) \cdot F\left(2a\right)\right\},$$

and subsequently identify $F\left(2a\right)$. All remaining identification arguments carry over again.

# 6 Conclusion

The dyadic link formation model considered in this paper allows for nonparametric homophily effects, unobserved individual fixed effects and unobservable idiosyncratic pairwise shocks that follow an unknown distribution. We establish nonparametric, semi-nonparametric and semiparametric identification results, which provide theoretical foundations for empirical analysis of many special cases of the model considered here.

Our identification results are rather complete. The findings characterize the identification (or lack thereof) of every unknown aspect of the model except for the exact realizations of idiosyncratic pairwise shocks, which are obviously unidentifiable. The identification of nuisance parameters, such as realizations of individual fixed effects and the distribution of idiosyncratic pairwise shocks, may assist in counterfactual analysis of the model, such as the evaluation of policy intervention in network formation. The results also provide an explicit and precise account of the identified sets, enabling a further determination of invariant properties within the identified sets and thereby clarifying economic interpretation of identified parameters.

The identification arguments provided in this paper are potentially useful for estimation and inference. Bonhomme, Lamadon, and Manresa (2017) propose a procedure for two-step estimation in a panel data setting that could potentially be adapted to our current problem. Specifically, they first use the "kmeans" algorithm to categorize discretized unobserved heterogeneity, and then construct a plug-in maximum likelihood estimator of the parameter of interest under fully parametric specifications. It remains unclear, however, whether their method can be fully adapted to the current network setting under nonparametric distribution of idiosyncratic shocks and nonparametric specification of homophily effects. If consistent and asymptotically normal estimators can indeed be constructed, then it will become feasible to conduct hypothesis testing on the magnitudes of homophily effects relative to degree het-erogeneity or idiosyncratic randomness, as the interquantile-range normalization avoids any a priori restriction on the scale of the homophily effect function and preserves the topological structures of the parameter space. As discussed in Graham (2017), such tests have potentially important empirical and policy implications. However, the construction of practical estimators and test statistics, as well as a characterization of their asymptotic behaviors, are not considered here and are left for future work.

As pointed out earlier, a limitation of the index link formation model considered here is that it does not explicitly incorporate link externality (or in other words, interdependent preferences), particularly in the plausible presence of *negative* link externality. A natural next step, of course, is to consider degree heterogeneity along with link externality. As discussed at least in Mele (2017a), it remains an open question whether degree heterogeneity can be separately identified from link externality in a single large network. We leave for future work a further investigation along this direction.

# Appendix

## A Alternative Assumptions

As discussed in Section 2 and shown in Section 4, Assumption 1 (degeneration of homophily function on the hyper-diagonal) is key to our (semi-)nonparametric identification results, while Assumption 5a (full dimensionality of the support of $W_{ij}$"), on the other hand, is necessary for the application of the identification results obtained under the nonparametric homophily setting to the parametric homophily setting. Below we present alternative assumptions corresponding to Assumption 1 and Assumption 5a, compare their relative strengths and discuss their relationship to the corresponding assumptions in Graham (2017), Candelaria (2016) and Toth (2017).

The original Assumptions 1 and 1a made in Section 2 are repeated below for easy reference. Recall that Assumption 1 is made for the nonparametric homophily setting while Assumption 1a is made for the parametric homophily setting under equation (4).

**Assumption. 1**(Degeneration of Homophily Function on the Hyper-Diagonal) $w : Supp\left(X_i\right) \times Supp\left(X_i\right) \to \mathbb{R}$ *is symmetric, continuous (with respect to the Euclidean norm on* $Supp\left(X_i\right) \times Supp\left(X_i\right) \subseteq \mathbb{R}^{2k}$*) and takes an unknown constant value* $\theta \in \mathbb{R}$ *on the hyper-diagonal in* $\mathbb{R}^k \times \mathbb{R}^k$*:* $\left\{\left(x_i, x_j\right) \in \mathbb{R}^k \times \mathbb{R}^k : \ x_i = x_j\right\}$.

**Assumption. 1a**(Component Interdependent Pseudo-Distance Function) Each component $\tilde{w}_m : \mathbb{R}^k \times \mathbb{R}^k \to \mathbb{R}$ of $\tilde{w} = \left(w_m\right)_{m=1}^k$ *is symmetric, continuous (with respect to the Euclidean norm) on* $\mathbb{R}^k$*, and* $\tilde{w}_m\left(x_i, x_j\right) = 0$ *whenever* $x_i = x_j$.

We first present Assumption 1' and 1'a, which are weaker than Assumption 1 and 1a, respectively: instead of requiring that the homophily effect function be degenerate or zero on the "hyper-diagonal", i.e., when two individuals share the same observable characteristics, Assumption 1' and 1'a only require that the homophily effect function be degenerate or zero on a "known curve" in the product space of two individuals' observable characteristics, which captures the underlying idea for our identification arguments.

**Assumption. 1'**(Degeneration of Homophily Function on a Known Hyper-Curve) $w : Supp\left(X_i\right) \times Supp\left(X_i\right) \to \mathbb{R}$ *is symmetric, continuous (with respect to the Euclidean norm on* $\mathbb{R}^k$*) and takes an unknown constant value* $\theta \in \mathbb{R}$ *on a known curve: there exists a known function* $g : Supp\left(X_i\right) \to Supp\left(X_i\right)$ *such that* $w\left(x_i, g\left(x_i\right)\right) \equiv \theta$ *for all* $x_i \in Supp\left(X_i\right)$.

**Assumption. 1'a**(Existence of a Curve of "Zero" Observable Pairwise Characteristics):

The known function $\tilde{w}$ is symmetric and continuous; moreover, $\forall x_i \in Supp(x_i)$ , there exists some $x_j \in Supp(X_i)$ such that $\tilde{w}(x_i, x_j) = \mathbf{0}$.[11]

Note that, the main identification arguments in Subsections 4.1-4.3 condition on pairs of individuals with observable characteristics that are known to produce degenerate homophily effects, which remains valid under Assumption 1' and 1'a. As long as such pairs of individuals are conditioned upon, the homophily effects become degenerate so that the homophily index $w(x_i, x_j)$ effectively becomes zero, and all subsequent identification arguments in Subsections 4.1-4.3 carry over with no change and establishes identification of $F$ and $A$.

In particular, notice that both Assumption 1' and 1'a require that the hyper-curve in the product space $Supp(X_i) \times Supp(X_i)$ of two individuals' observable characteristics pass through every possible value of $Supp(X_i)$ , so that we can apply our identification arguments to all individuals and recover all their fixed effects. Then, following the arguments in Subsections 4.4 and 4.5, we may identity the nonparametric homophily effect function $w$ or parametric homophily effect parameters $\beta_0$.

Hence, Assumption 1' and 1'a are technically the weaker versions of assumptions we need to obtain our identification results, and they more clearly demonstrate the underlying idea of and the required condition for our identification arguments. However, as our main model is motivated as a link formation model featured with homophily effect, we proceed in the main text the stronger Assumption 1 and 1a that emphasize the degeneracy of the function $w$ when individuals have completely the same observable characteristics ($x_i = x_j$), as they offer a clearer economic interpretation of $w$ as homophily effects, the tendency of individuals to link with individuals of *similar* ($x_i \approx x_j$) characteristics.

For the parametric homophily setting $w(x_i, x_j) = \tilde{w}(x_i, x_j)' \beta_0$, we may further strengthen Assumption 1a to Assumption 1b and 1c below, which allows an even more clearer interpretation of homophily effects based on (pseudo-)metrics on the product space of two individuals' observable characteristics. In particular, note that Assumption 1c essentially summarizes the corresponding restrictions made in Toth (2017, Assumptions 1, 2).

**Assumption. 1b**(Component Interdependent Distance Function) Each component $\tilde{w}_m : \mathbb{R}^k \times \mathbb{R}^k \to \mathbb{R}$ of $\tilde{w} = (w_m)_{m=1}^k$ *is symmetric, continuous with respect to the Euclidean norm on* $\mathbb{R}^k$, *and is more over a (pseudo-)metric on* $\mathbb{R}^{2k}$*: : it is nonnegative, symmetric, and* $\tilde{w}_m(x_i, x_j) = 0$ *if (and only if)* $x_i = x_j$.

**Assumption. 1c**(Component Separable Distance Function): $\tilde{w}(x_i, x_j) = (\tilde{w}_m(x_{im}, x_{jm}))_{m=1}^k$

[11]Alternatively, stated in terms of the pairwise observable characteristics $\tilde{W}_{ij}$, we basically require that, in the population distribution, $\forall i$, there exists some $j$ such that $\tilde{W}_{ij} = \mathbf{0}$.

*for* $(x_i, x_j) \in \mathbb{R}^k \times \mathbb{R}^k$*, whose each component* $w_m$ *(*$m = 1, ..., k$*) is continuous with respect to the Euclidean norm on* $\mathbb{R}^2$*, and it is moreover a (pseudo-)metric on* $\mathbb{R}^2$ *: it is nonnegative, symmetric, and* $\tilde{w}_m(x_{im}, x_{jm}) = 0$ *if (and only if)* $x_{im} = x_{jm}$*.*

The only difference between Assumption 1b and 1c lies in whether we allow the (pseudo)-metric functions to be contingent on the whole vector of observable characteristics or to be separably defined on each component of observable characteristics. Assumption 1b and 1c are not equivalent even in consideration of basis transformation of coordinates, as the function $\tilde{w}$ is allowed to be nonlinear.

We now briefly discuss an alternative assumption to Assumption 5a, which is repeated below for easier reference:

**Assumption. 5a**(Full-Dimensional Support of $\tilde{W}_{ij}$): *Supp* $\left(\tilde{W}_{ij}\right) =$ *Supp* $(\tilde{w}(X_i, X_j))$ *contains* $k$ *linearly independent nonzero vectors* $\overline{W}_m \in \mathbb{R}^k$*.*

Specifically, Assumption 5b below, jointly with Assumption 1b above, jointly imply Assumption 5a and 1a, so we could regard Assumption 5b as a roughly stronger condition to ensure full dimensionality.

**Assumption. 5b**(Support of $X_i$ with Nonempty Interior): *Supp* $(X_i)$ *has nonempty interior in* $\mathbb{R}^k$*.*

Assumption 5b is familiar in the literature as a condition to ensure full dimensionality. However, notice that Assumption 5b in particular essentially rules out discrete observable characteristics, such as ethnicity and gender, which often times are the most basic individual characteristics to include in the link formation models considered here. We emphasize that under the working sets of assumptions in the main text, arbitrary (continuous, discrete and mixed) forms of observable characteristics can be accommodated.

We now summarize the relative strengths of the assumptions discussed above by the following lemma.

**Lemma 2.**

*(i) A1b* $\Rightarrow$ *A1a* $\Rightarrow$ *A1'a* $\Rightarrow$*A1'.*

*(ii) A1a* $\Rightarrow$ *A1* $\Rightarrow$ *A1'.*

*(iii) (A1b* $\wedge$ *A5b)* $\Rightarrow$ *(A1a* $\wedge$ *A5a).*

## Proof of Lemma 2

*Proof.* (i) and (ii) are obvious. For (iii), Assumption 1a follows immediately from Assumption 1b. We now prove Assumption 5a under Assumptions 1b and 5b.

As $Supp\left(X_i\right)$ has nonempty interior, it contains a closed ball $\overline{B\left(\overline{x}, \overline{\epsilon}\right)} \subseteq int\left(Supp\left(X_i\right)\right)$ centered at $\overline{x}$ with radius $\overline{\epsilon} > 0$ (inside an open ball with larger radius in $int\left(Supp\left(X_i\right)\right) \subseteq \mathbb{R}^k$). Let

$$\overline{x}^{(m)} := \overline{x} + \overline{\epsilon} e_m,$$
$$\overline{W}_m := w_m\left(\overline{x}, \overline{x}^{(m)}\right) e_m$$

Then $\overline{x}^{(m)} \in \overline{B\left(\overline{x}, \overline{\epsilon}\right)} \in int\left(Supp\left(X_i\right)\right)$ and $\left\{\overline{W}_m : m = 1, ..., k\right\}$ are linearly independent. This concludes the proof for Assumption 5a. □

# B Discussion of "Normalization Without Loss of Generality"

In this section, we provide a formal analysis of "normalization without loss of generality". We now provide an overview of the line of reasoning before presenting the formal definitions and analysis starting from Appendix B.1.

In Section 3, we have argued that all relevant model specifications and assumptions are invariant under the transformation $\psi_{a,b,c}$ defined in equation (5), we thus call $\psi_{a,b,c}$ a *modeling equivalent transformation* of the unknown $(w, A, U)$. We say that two values of the unknown $(w, A, U)$ and $\left(\hat{w}, \hat{A}, \hat{U}\right)$ are *modeling equivalent*, written as $(w, A, U) \sim \left(\hat{w}, \hat{A}, \hat{U}\right)$, if there exists some $(a, b, c) \in \mathbb{R} \times \mathbb{R} \times \mathbb{R}_{++}$ such that $\psi_{a,b,c}(w, A, U) = \left(\hat{w}, \hat{A}, \hat{U}\right)$, i.e., if there exists a (bijective) modeling equivalent transformation in the form of $\psi_{a,b,c}$ between them. It is then straightforward to check that "$\sim$" is indeed an equivalence relation: it is reflexive, symmetric and transitive.[12] We call "$\sim$" the *modeling equivalence relation* on the space of the unknown $(w, A, U)$ under our model, which then induces a quotient space, or a *modeling equivalence class partition*, of the space of $(w, A, U)$.

Intuitively, and as rigorously shown in the following subsections, no proper nonempty subsets of any cell of the modeling equivalence partition can be identified, as nothing in the model (or the data) can differentiate any two points that lie in the same modeling equivalence class. In particular, we cannot tell the true value of $(w, A, U)$ and the corresponding $F^{-1}$ (the inverse CDF of $U_{ij}$) from their positive affine transformations in the form of (5). Hence, given any counterfactual parameter or any economic question, which can be essentially regarded as a function of the unknown, it is identified if and only if it is measurable with respect to the sigma algebra generated by the modeling equivalence class partition, or in other words, if and only if it is invariant under the modeling equivalent transformations defined here.

[12] The requirement of bijectivity in the definition of the modeling equivalent transformation $\psi_{a,b,c}$ is key to ensure symmetry.

We define *normalization* as an arbitrary way of labeling each modeling equivalence class by one of its element. In other words, by normalization we represent each modeling equivalence class with one point, so that we may obtain results in the form of point identification, but in the end the "identified point" is fundamentally an element of the quotient space, which stands for an equivalence class of the unknown objects in the model. This implies, by Theorem 4 in Appendix B.3, that normalization defined in this sense is *without loss of generality* with respect to the econometric model's ability to identify unknown aspects of the model, to inform about counterfactual parameters as well as to provide answers to economic questions.

## B.1 Notations

To abstract from the details of the main model that are irrelevant in this section, we simplify the notations for the elements in model (3) as follows. We rewrite model (3) along with all specifications and assumptions (Assumptions 1, 2, 3 and 4) as

$$f(\mathbf{X}, \mathbf{U}, \gamma_0) = 0, \tag{17}$$

where:

- $\mathbf{X} := (D, X)$ collects all observable random elements, regarded as measurable functions from some underlying probability space $(\Omega, \mathscr{F}, \mathbb{P})$ to the corresponding range spaces. For example, we understand $D$ as the vector of measurable functions $D_{ij} : (\Omega, \mathscr{F}) \to \left(\{0, 1\}, 2^{\{0,1\}}\right)$ for all $i \neq j$.

- $\mathbf{U} := (A, U)$ collects all unobservable random elements, again regarded as measurable functions from the underlying probability space to the corresponding range spaces. For example, we understand $U$ as the vector of measurable functions $U_{ij} : (\Omega, \mathscr{F}) \to (\mathbb{R}, \mathscr{B}(\mathbb{R}))$ for all $i \neq j$.

- $\gamma_0 := (w, F)$ collects the true values of all unknown but fixed potentially infinite-dimensional parameters. Note that $F$, the unknown CDF of $U_{ij}$, is innocuously redundant here: $F$ is determined implicitly by $\mathbf{U}$, but we write out $F$ explicitly to emphasize its role as an infinite-dimensional parameter of our model.

- $f$ is a known function that represents the "structural equation" (3) along with all other model specifications and assumptions (Assumptions 1, 2, 3 and 4):

$$f(\mathbf{X}, \mathbf{U}, \gamma_0) := 1 - \prod_{i \neq j} \mathbf{1}\{D_{ij} = \mathbf{1}\{w(X_i, X_j) + A_i + A_j \geq U_{ij}\}\}$$
$$\cdot \mathbf{1}\{\text{All other relevant model specifications hold}\}$$
$$\cdot \mathbf{1}\{\text{Assumptions 1, 2, 3 and 4 hold}\}.$$

  Note that the arguments $\mathbf{X}, \mathbf{U}$ of $f$ are random variables (measurable functions), so $f$ is a real-valued (binary) functional on the space of all possible $(\mathbf{X}, \mathbf{U}, \gamma)$. For example, the IID assumption "$U_{ij} \sim_{iid} F$" can be readily incorporated into $f$.

The tuple $(\mathbf{X}, \mathbf{U}, \gamma_0, f)$, along with their corresponding domains and range spaces, summarizes all relevant elements of our main model. In particular, notice that the known function $f$ incorporates not only the structural equation but also all the other relevant model specifications and assumptions. In fact, mathematically there are no fundamental distinctions between model specifications and model assumptions: they are all integral parts of the model written down by an econometrician.

Even though the elements $(\mathbf{X}, \mathbf{U}, \gamma_0, f)$ above are defined for the particular model considered in this paper, a further reflection suggests that the decomposition of an econometric model into four components, where $\mathbf{X}$ stands for the observable "data", $\mathbf{U}$ for the unobservable "error", $\gamma_0$ for the unknown "parameter" and $f$ for the known "assumptions", is valid much more generally. Hence, much of the proposed definitions and results in subsequent Subsections may be applied in more general settings. Yet for concreteness, we maintain the definitions of $(\mathbf{X}, \mathbf{U}, \gamma_0, f)$ for the main model presented in Section 2.

## B.2 Modeling Equivalence and Normalization

In this Subsection, we propose a formal definition of normalization. In particular, we present the concept of modeling equivalence, and define normalization as a particular way of labeling modeling equivalence classes.

In the rest of this section, we write $(\mathcal{U}, \Gamma)$ to denote the space of all admissible values of the unknown elements $(\mathbf{U}, \gamma)$ in the model, and call the product space $\mathcal{U} \times \Gamma$ the "space of the unknown". We then define the modeling equivalence relation on the space of the unknown.

**Definition 2** (Modeling Equivalence)**.** Let $(\mathbf{U}, \gamma)$ and $(\mathbf{U}', \gamma')$ be two points in $\mathcal{U} \times \Gamma$. We say that $(\mathbf{U}, \gamma)$ and $(\mathbf{U}', \gamma')$ are ***modeling equivalent***, written as $(\mathbf{U}, \gamma) \sim (\mathbf{U}', \gamma')$, if there exists a bijective transformation on the space of the unknown $\psi : \mathcal{U} \times \Gamma \longmapsto \mathcal{U} \times \Gamma$ such that

$$f\left(X, \psi\left(\mathbf{U}, \gamma\right)\right) = f\left(X, \mathbf{U}, \gamma\right), \quad \forall \left(\mathbf{U}, \gamma\right) \in \mathcal{U} \times \Gamma. \tag{18}$$

Recall that, in this paper, $\psi$ takes the form of the positive affine transformation (5) defined in Section 3.

**Lemma 3** (Modeling Equivalence Relation)**.** *"$\sim$" is an* ***equivalence relation*** *on* $\mathcal{U} \times \Gamma$.

*Proof.* Reflexivity and transitivity are trivial. Symmetry follows from the bijectivity of $\psi$. □

**Definition 3** (Modeling Equivalence Class)**.**

- We denote the quotient space of $\mathcal{U} \times \Gamma$ with respect to the equivalence relation "$\sim$" by

$$P_{\sim} := \left(\mathcal{U} \times \Gamma\right) / \sim .$$

- We refer to $P_{\sim}$ as the ***modeling equivalence partition*** of $\mathcal{U} \times \Gamma$.
- We refer to each element $p \in P_{\sim}$, a ***cell*** of the modeling equivalence partition $P_{\sim}$, as a ***modeling equivalence class*** of $\mathcal{U} \times \Gamma$.
- We write $p\left(\mathbf{U}, \gamma\right) \in P_{\sim}$ to denote the modeling equivalence class that $\left(\mathbf{U}, \gamma\right)$ lies in.

We now provide a formal discussion about modeling equivalence and *observational equivalence*, a widely used concept in the literature. Specifically, observational equivalence is defined only with respect to the data but not necessarily with respect to modeling specifications and assumptions, so we need to augment the definition of equivalence transformations in the following way.

**Definition 4** (Observational Equivalence)**.** Consider two specifications of $(f, \mathbf{U}, \gamma)$ and $\left(f^{'}, \mathbf{U}^{'}, \gamma^{'}\right)$. We say that $(f, \mathbf{U}, \gamma)$ and $\left(f^{'}, \mathbf{U}^{'}, \gamma^{'}\right)$ are ***observationally equivalent***, written as $(f, \mathbf{U}, \gamma) \sim_{obs} \left(f^{'}, \mathbf{U}^{'}, \gamma^{'}\right)$, if there exists a bijective transformation on the space of the unknown $\psi : \mathcal{U} \times \Gamma \longmapsto \mathcal{U}^{'} \times \Gamma^{'}$ as well as a bijection $\phi$ that maps between the space of $f$ and the space of $f^{'}$ such that

$$f^{'}\left(X, \mathbf{U}^{'}, \gamma^{'}\right) \equiv \left[\phi\left(f\right)\right]\left(X, \psi\left(\mathbf{U}, \gamma\right)\right) \equiv f\left(X, \mathbf{U}, \gamma\right). \tag{19}$$

The key difference between *modeling equivalence* and *observational equivalence* lies in whether we allow a transformation of $f$ to $f^{'}$, which encode all modeling specifications and assumptions.

We now present the following proposition, which states that modeling equivalence is a strict refinement of observational equivalence.

**Proposition 2.** *Modeling equivalence implies observational equivalence; the converse is not true in general.*

The first half follows trivially from the definitions. The second half can be seen from Example 1 in Section 3.

**Definition 5** (Normalization)**.** We say a transformation $\psi_N$ of the unknown is a ***normalization*** if

$$\begin{aligned}(\mathbf{U},\gamma)\sim\left(\mathbf{U}^{'},\gamma^{'}\right)&\Rightarrow\psi_N(\mathbf{U},\gamma)=\psi_N\left(\mathbf{U}^{'},\gamma^{'}\right)\sim(\mathbf{U},\gamma)\\(\mathbf{U},\gamma)\nsim\left(\mathbf{U}^{'},\gamma^{'}\right)&\Rightarrow\psi_N(\mathbf{U},\gamma)\nsim\psi_N\left(\mathbf{U}^{'},\gamma^{'}\right)\end{aligned}$$

i.e. $\psi_N$ maps different $(\mathbf{U},\gamma)$ within a modeling equivalence class of $\mathcal{U}\times\Gamma$ to a *single element in the same equivalent class*. Equivalently, a normalization is a "labeling" of each modeling equivalent class for selecting exactly one element from that class.

Note that in contrast to the definition of modeling equivalence, there is no requirement for $\psi$ to be bijective. Actually, it should not be, as this mapping labels all elements within a given equivalent class to a single element in that class.

Under the nonparametric homophily setting, as proposed in Section 3, we normalize the location of $\theta_0 := w(\overline{x},\overline{x})$, the $\alpha$-th quantile of $U_{ij}$, and the interquantile range between the $\alpha$-th and the $(1-\alpha)$-th quantiles of $U_{ij}$. This normalization operation is characterized by the transformation $\psi_N : \mathcal{U}\times\Gamma\longmapsto\mathcal{U}\times\Gamma$ defined below:

$$\psi_N(\mathbf{U},\gamma):=\left(\frac{1}{c}\left(w-w(\overline{x},\overline{x})\right),\frac{1}{c}\left(A-F^{-1}(\alpha)\right),\frac{1}{c}\left(U-w(\overline{x},\overline{x})-2F^{-1}(\alpha)\right)\right)$$

where $\mathbf{U}\equiv(A,U)$, $\gamma\equiv(w,F)$ and $c:=F^{-1}(1-\alpha)-F^{-1}(\alpha)$, which selects a single representative element from each modeling equivalent class in a systematic way. Specifically, given an equivalence class $p\in P_\sim$, $\psi_N$ selects the unique $(\mathbf{U}_p,\gamma_p)\equiv(A_p,U_p,w_p,F_p)$ in $p$ such that $w_p(\overline{x},\overline{x})=0$, $F_p^{-1}(\alpha)=0$ and $F_p^{-1}(1-\alpha)=1$.

## B.3 "Without Loss of Generality"

We now explain why the normalization defined above, as a labeling of modeling equivalence classes, is without loss of generality in a formal and precise sense. Roughly speaking, we define "without loss of generality" as an invariant property of any counterfactual parameters of the model under arbitrary alternative normalization. For expositional simplicity, we focus on real scalar counterfactual parameters below. However, the arguments can be extended to higher-dimensional counterfactual.

**Definition 6** (Counterfactual Parameter)**.** A *counterfactual (question)* is a known function $q : \mathcal{U} \times \Gamma \to \mathbb{R}$. We refer to $q(\mathbf{U}_0, \gamma_0)$ as a **counterfactual parameter** (an answer to the question), where $(\mathbf{U}_0, \gamma_0)$ denote the true values of the unknown aspects of the model.

Counterfactual parameters are conceptually important, as they incorporate meaningful (economic) questions about the unknown aspects of the model, which basically constitutes the ultimate goal of an econometrician that works with a particular econometric model.

**Lemma 4** (Limit of Identification)**.** *If a counterfactual $q : \mathcal{U} \times \Gamma \to \mathbb{R}$ is not measurable with respect to the sigma-algebra generated by the modeling equivalence partition of $\mathcal{U} \times \Gamma$, then there exists some true values of the unknown $(\mathbf{U}_0, \gamma_0)$ such that the counterfactual parameter $q(\mathbf{U}_0, \gamma_0)$ cannot be point identified.*

*Proof.* If $q : \mathcal{U} \times \Gamma \to \mathbb{R}$ is not measurable with respect to the modeling equivalence class partition $P_\sim$ of $\mathcal{U} \times \Gamma$, then there exists some $p \in P_\sim$ such that $q$ is not constant on $p$, i.e., there exists two distinct values $(\mathbf{U}, \gamma) \sim (\mathbf{U}', \gamma') \in p$ such that

$$q(\mathbf{U}, \gamma) \neq q\left(\mathbf{U}', \gamma'\right). \tag{20}$$

Now suppose the true values $(\mathbf{U}_0, \gamma_0) \in p$. Then $(\mathbf{U}_0, \gamma_0) \sim (\mathbf{U}, \gamma) \sim (\mathbf{U}', \gamma')$, so there exist some $a, b, a', b' \in \mathbb{R}$ and $c, c' > 0$ such that

$$(\mathbf{U}_0, \gamma_0) = \psi_{a,b,c}(\mathbf{U}, \gamma) = \psi_{a',b',c'}\left(\mathbf{U}', \gamma'\right)$$

where $\psi_{a,b,c}$ and $\psi_{a',b',c'}$ are defined by (5) in Section 3. Moreover, by the definition of "$\sim$", we have

$$f\left(\mathbf{X}, \psi_{a,b,c}\left(\tilde{\mathbf{U}}, \tilde{\gamma}\right)\right) = f\left(\mathbf{X}, \tilde{\mathbf{U}}, \tilde{\gamma}\right), \quad \forall \tilde{\mathbf{U}}, \tilde{\gamma}. \tag{21}$$

Now we define a new model specification $\overline{f}$ by

$$\overline{f}\left(\mathbf{X}, \tilde{\mathbf{U}}, \tilde{\gamma}\right) := f\left(X, \psi_{a,b,c}\left(\tilde{\mathbf{U}}, \tilde{\gamma}\right)\right),$$

$$\underline{f}\left(\mathbf{X}, \tilde{\mathbf{U}}, \tilde{\gamma}\right) := f\left(X, \psi_{a',b',c'}\left(\tilde{\mathbf{U}}, \tilde{\gamma}\right)\right), \quad \forall \tilde{\mathbf{U}}, \tilde{\gamma}.$$

In the redefined model $\left(\mathbf{X}, \tilde{\mathbf{U}}, \tilde{\gamma}, \overline{f}\right)$, the *new true value of the unknown* is clearly given by

$$\psi^{-1}_{a,b,c}\left(\mathbf{U}_0, \gamma_0\right) = \left(\mathbf{U}, \gamma\right),$$

which is directly imported from $(\mathbf{X}, \mathbf{U}_0, \gamma_0, f)$ by the construction of$\overline{f}$. Similarly, for the $\left(\mathbf{X}, \tilde{\mathbf{U}}, \tilde{\gamma}, \underline{f}\right)$, the true value of the unknown is given by $\left(\mathbf{U}', \gamma'\right)$.

However, by (21), the constructions of $\overline{f}$ and $\underline{f}$ are in the meanwhile degenerate constructions:

$$\overline{f}\left(\mathbf{X}, \tilde{\mathbf{U}}, \tilde{\gamma}\right) = f\left(\mathbf{X}, \psi_{a,b,c}\left(\tilde{\mathbf{U}}, \tilde{\gamma}\right)\right) = f\left(\mathbf{X}, \tilde{\mathbf{U}}, \tilde{\gamma}\right), \quad \forall \tilde{\mathbf{U}}, \tilde{\gamma}$$

implies that $\overline{f} \equiv f$. By a similar argument, we conclude that

$$f \equiv \overline{f} \equiv \underline{f}. \tag{22}$$

Let $q_{ID}\left(\mathbf{X}, f\right)$ denote a generic identified set of the counterfactual parameter $q\left(\mathbf{U}_0, \gamma_0\right)$ obtained in any given valid identification result for the model $(\mathbf{X}, \mathbf{U}_0, \gamma_0, f)$. Clearly, as $(\mathbf{U}_0, \gamma_0)$ are unobservable or unknown, they cannot be part of the arguments of $q_{ID}(\cdot)$: any identification results, obtained by any econometrician for any model, are ultimately based on the observable or known aspects of the model summarized by $(\mathbf{X}, f)$. Our goal is to show that $q\left(\mathbf{U}_0, \gamma_0\right)$ cannot be point identified.

First, notice that by (22), we can immediately deduce that

$$q_{ID}\left(\mathbf{X}, f\right) \equiv q_{ID}\left(\mathbf{X}, \overline{f}\right) \equiv q_{ID}\left(\mathbf{X}, \underline{f}\right). \tag{23}$$

As $f \equiv \overline{f} \equiv \underline{f}$ across the "three" models, and the observable data $\mathbf{X}$ are also the same across the "three models", the "three" models, from the perspective of an econometrician, are just the same single model, so the identification results must be exactly the same.

Second, provided that $q_{ID}\left(\mathbf{X}, f\right)$ is a valid identification result, it must be that

$$q\left(\mathbf{U}_{\mathbf{0}}, \gamma_0\right) \in q_{ID}\left(\mathbf{X}, f\right). \tag{24}$$

Similarly, as $(\mathbf{U}, \gamma)$ and $\left(\mathbf{U}', \gamma'\right)$ are the true values under $\left(\mathbf{X}, \overline{f}\right)$ and $\left(\mathbf{X}, \underline{f}\right)$ respectively, we also have

$$q\left(\mathbf{U}, \gamma\right) \in q_{ID}\left(\mathbf{X}, f\right), \quad q\left(\mathbf{U}', \gamma'\right) \in q_{ID}\left(\mathbf{X}, f\right) \tag{25}$$

Third, (23), (24) and (25), we deduce that

$$\left\{q\left(\mathbf{U}_0,\gamma_0\right),q\left(\mathbf{U},\gamma\right),q\left(\mathbf{U}^{'},\gamma^{'}\right)\right\}\subseteq q_{ID}\left(\mathbf{X},f\right)$$

which, together with (20), implies

$$\#q_{ID}\left(\mathbf{X},f\right)\geq\#\left\{q\left(\mathbf{U},\gamma\right),q\left(\mathbf{U}^{'},\gamma^{'}\right)\right\}=2.$$

This proves that $q\left(\mathbf{U}_0,\gamma_0\right)$ cannot be point identified. □

Lemma 4 implies that, if the answer to an economic question about the unknown aspects of the model is not invariant on every given modeling equivalence class, then the answer to this question is unidentified under some true values of the unknown.

**Theorem 4** (Normalization WLOG)**.** *A normalization (as defined in this paper) is* ***without loss of generality (WLOG)*** *in the following sense: Let $q$ be any counterfactual and let $\psi_N$ be a normalization. Then:*

- *(i) The (point) identifiability of $q\left(\mathbf{U}_0,\gamma_0\right)$ is invariant to the choice of $\psi_N$.*
- *(ii) In the case where $q\left(\mathbf{U}_0,\gamma_0\right)$ is indeed point identified, the identified value $q\left(\mathbf{U}_0,\gamma_0\right)$ is also invariant to the choice of $\psi_N$.*

*Proof.* Note that (ii) is trivial. For (i), it is conceptually clear that the identifiability of $q\left(\mathbf{U}_0,\gamma_0\right)$ is determined by, as suggested in Lemma 4, the modeling equivalence classes of the model. Normalization, as a labeling of the equivalence classes, should thus be irrelevant. For rigor of exposition, we present below a mathematical argument why normalization is irrelevant.

After imposing the normalization $\psi_N$, the model is fundamentally transformed into

$$\begin{aligned}f_N\left(\mathbf{X},\tilde{\mathbf{U}},\tilde{\gamma}\right)&:=f\left(\mathbf{X},\psi_N\left(\tilde{\mathbf{U}},\tilde{\gamma}\right)\right)\\&=f\left(\mathbf{X},\psi_{a,b,c}\left(\tilde{\mathbf{U}},\tilde{\gamma}\right)\right),\text{ for some }a,b\in\mathbb{R},c>0\\&=f\left(\mathbf{X},\tilde{\mathbf{U}},\tilde{\gamma}\right),\quad\forall\tilde{\mathbf{U}},\tilde{\gamma}\end{aligned}$$

which is again a fundamentally degenerate redefinition, i.e., $f_N\equiv f$. Hence, any valid identification result derived after the normalization must coincide exactly as a corresponding valid identification result for the true (unnormalized) model, i.e.,

$$q_{ID}\left(\mathbf{X},f_N\right)=q_{ID}\left(\mathbf{X},f\right).$$

This establishes the invariance of any valid identification results to the normalization transformation $f_N$. □

We emphasize that "without loss of generality" does not mean "powerful": it simply means that normalization leads to *no change* at all in the generality and the ability of the model to inform about counterfactual, regardless of the exact level of generality of the model itself. Normalization, by its name, should be and indeed is irrelevant: in this paper, it simply serves as one among many ways to help characterize the identified set, but it has no effects whatsoever on the identification results per se.

It is also crucial that, the "point" identification result under normalization is translated back as set identification results in recognition of the fundamental lack of point identifiability in the binary response model considered here. Though this seems rather redundant, the set identification results, as presented in Theorems 1 and 2, provide a precise characterization of the structures of the modeling equivalence classes, thus allowing a precise analysis of what counterfactual parameters, are point identified and what are not identified. For example, the invariant properties under the positive affine transformations defined in 1 and 2 are identified.

If, however the identification results are presented as "point identification under normalization", while the structure of normalization is kept implicit, then the information on the invariant properties within the identified modeling equivalent class becomes implicit and thus easily overlooked. This problem becomes especially salient when empirical researchers conduct counterfactual analysis on normalized models and make certain claims on quantities whose interpretations are not invariant under alternative normalization. For example, in the empirical industrial organization literature that study differentiated products, positive affine transformation via location and scale normalization similar to the normalization considered in this paper is often times imposed. In this case, counterfactual parameters in the form of ratios (say, percentage changes in consumer welfare), may not be identified, as a shift in location along with a change in scale may lead to wild changes in ratios, potentially obfuscating a clear interpretation of ratios.

# C Nonlinear Coupling of Fixed Effects

In this section, we relax the specification that individual fixed effects are aggregated in a linear form: $A_i + A_j$, and considers potentially nonlinear coupling of $A_i$ and $A_j$ into $\phi(A_i, A_j)$, which allows more flexible specification of how $A_i$ and $A_j$ are coupled into an index term.

As well-known in the literature on identification, it turns out that the nonlinearity of $\phi$

may help with identification, as the curvature of $\phi$ may provide richer information about the scale or location parameters, which in the linear additive case are guaranteed to be unidentifiable. Hence, Theorems 1 and 2, along with their proofs in previous sections under linear aggregation of fixed effects, may be regarded as results for the "worst-case scenario".

In Subsection C.1, we present the formal specification of nonlinearly coupled fixed effects, provide some general discussion on how the identification arguments in previous sections can be adapted. In Subsection C.2, we solve out in details a concrete illustrating example that conveys the main insights. Subsections C.3 and C.4 then provide further characterizations of two general forms of sufficient conditions for *unidentifiability* of location and scale.

## C.1 Adaptions for the Nonlinearity of $\phi$

We now revisit model (9) in Remark 2:

$$D_{ij} = \left\{ w\left(X_i, X_j\right) + \phi\left(A_i, A_j\right) \geq U_{ij} \right\}.$$

where $w$ is nonparametrically specified and $\phi : \mathbb{R}^2 \to \mathbb{R}$ is a *known* potentially nonlinear function on which we impose the following assumption:

**Assumption. 6** (Nice Coupling Function) $\phi : \mathbb{R}^2 \to \mathbb{R}$ *is a symmetric, differentiable, strictly increasing (in both arguments) and surjective (in both arguments) function.*

Assumption 6 mainly ensures that the coupled fixed effects are still "smoothly ordered" with a "full support": continuity rules out jumps in the coupled effects of $A_i$ and $A_j$ on linking probabilities, strict monotonicity preserves the implications of individual fixed effects on relative popularity, and surjectivity implies that given an arbitrary $A_i$, the other individual's fixed effect $A_j$ is still pivotal in the determination of the coupled effect.

As discussed in Remark 2, setting the two quantiles $Q_\alpha, Q_{1-\alpha}$ to $0, 1$ are no longer valid normalization. However, the identification arguments can be adapted to accommodate the lack of normalization. We now simply write

$$Q_\alpha \left[U_{ij}\right] = q_\alpha, \; Q_{1-\alpha} \left[U_{ij}\right] = q_{1-\alpha}$$

for $\alpha \in \left(0, \frac{1}{2}\right)$ and two *unknown* constants $q_\alpha < q_{1-\alpha}$. We treat $q_\alpha, q_{1-\alpha}$ effectively as location and scale parameters that we seek to identify.

We now carry out the arguments in Subsections 4.1-4.3 with some minor technical adaptions and a major change of interpretation, which we discuss more specifically below.

(a) Subsection 4.1 (identification of individuals with same observable characteristics and same fixed effects): the arguments in this Subsection carry over without any need for changes.

(b) Subsection 4.2 (identification of individuals with same observable characteristics and certain fixed effects): the arguments in this Subsection carry over almost without change. The only difference lies in that the two initially "identified" fixed effects must now be written as

$$\overline{\phi}^{-1}(q_\alpha), \overline{\phi}^{-1}(q_{1-\alpha}),$$

where $\overline{\phi}: \mathbb{R} \to \mathbb{R}$, defined by $\overline{\phi}(a) := \phi(a, a)$ for all $a \in \mathbb{R}$, is a *known*, potentially nonlinear, symmetric, continuous, strictly increasing and surjective function, thus admitting an inverse function $\overline{\phi}^{-1}$. With $q_\alpha, q_{1-\alpha}$ unknown, the two initial identified levels of fixed effects $\overline{\phi}^{-1}(q_\alpha)$ and $\overline{\phi}^{-1}(q_{1-\alpha})$ are also unknown. Yet the subtlety here is that, even though the two numbers $\overline{\phi}^{-1}(q_\alpha), \overline{\phi}^{-1}(q_{1-\alpha})$ are unknown, we can nevertheless identify pairs of individuals who share the same unknown fixed effects $\overline{\phi}^{-1}(q_\alpha)$ or $\overline{\phi}^{-1}(q_{1-\alpha})$. More precisely, what we identify here are the indicator functions

$$\mathbf{1}\left\{A_i = A_j = \overline{\phi}^{-1}(q_\alpha)\right\}, \ \mathbf{1}\left\{A_i = A_j = \overline{\phi}^{-1}(q_{1-\alpha})\right\},$$

but not $\overline{\phi}^{-1}(q_\alpha)$ and $\overline{\phi}^{-1}(q_{1-\alpha})$ themselves.

(c) Subsection 4.3 (identification of the CDF $F$ and the fixed effects $A$): we might then adapt the "in-fill" and "out-expansion" arguments in this Subsection to extend the identification results to the whole real line. Specifically, for the in-fill algorithm, instead of considering bisection points, we now consider recursive iterations of the following identification equations:

$$F(\phi(a_i, a_j)) = \mathbb{E}\left[D_{ij} \middle| x_i = x_j = \overline{x},\ A_i = a_i, A_j = a_j\right],$$

$$\mathbf{1}\left\{A_h = A_k = \overline{\phi}^{-1}(\phi(a_i, a_j))\right\} = \mathbf{1}\left\{\mathbb{E}\left[D_{hk} \middle| x_h = x_k = \overline{x},\ A_h = A_k\right] = F(\phi(a_i, a_j))\right\},$$

The first equation extends current knowledge of individuals with certain fixed effects $A_i = a_i$ and $A_j = a_j$ to the identification of $F$ at a new point $\phi(a_i, a_j)$, which is unknown but nevertheless characterizable by the two initial unknown quantiles $q_\alpha, q_{1-\alpha}$. The second equation then utilizes the knowledge of $F(\phi(A_i, A_j))$, a known number in $[0, 1]$, to back out individuals that share a new unknown level of fixed effects

$\overline{\phi}^{-1}(\phi(a_i, a_j))$. It is straightforward to check that

$$\overline{\phi}^{-1}(\phi(a_i, a_j)) \in (\min\{a_i, a_j\}, \max\{a_i, a_j\}),$$

so this process still in-fills the interval $\left[\overline{\phi}^{-1}(q_\alpha), \overline{\phi}^{-1}(q_{1-\alpha})\right]$ for $A$ (and correspondingly for $F$) inductively. Similarly, an adapted "out-expansion" algorithm help extend the identification results from $\left[\overline{\phi}^{-1}(q_\alpha), \overline{\phi}^{-1}(q_{1-\alpha})\right]$ to $\mathbb{R}$. The problem, however, is that even though the identification result is known to be extended to the whole real line $\mathbb{R}$, we have not yet established identification at each particular point, say, $0 \in \mathbb{R}$. Again the subtlety here lies in that every identification result obtained so far is ultimately characterized with respect to the two *unknown* numbers $q_\alpha$ and $q_{1-\alpha}$, so we only know the values of $F$ and $A$ on points in $\mathbb{R}$ that are described as known functions of $q_\alpha$ and $q_{1-\alpha}$. For example, for $a = \overline{\phi}^{-1}\left(\phi\left(\overline{\phi}^{-1}(q_\alpha), \overline{\phi}^{-1}(q_{1-\alpha})\right)\right)$, we can identify

$$F(a) = \mathbb{E}\left[D_{ij} | x_i = x_j = \overline{x},\ A_i = \overline{\phi}^{-1}(q_\alpha), A_j = \overline{\phi}^{-1}(q_{1-\alpha})\right]$$

as a known number in $(0, 1)$, even though we do not know the exact value of $a$ as a real number. As a metaphor, we could think of the whole real line $\mathbb{R}$ as "encrypted" by the two unknown "encryption keys" $q_\alpha$ and $q_{1-\alpha}$. As a result, the main remaining question is whether we can identify $q_\alpha$ and $q_{1-\alpha}$ themselves, and then use them as keys to "decrypt" previous identification results on the real line.

We now proceed to discuss the identification (or lack thereof) of $q_\alpha$ and $q_{1-\alpha}$.

First, notice that, if $\phi$ is linear, i.e., $\phi(a, b) = a + b$, then the model is obviously reduced back to our original setting in Sections 2-4 and the identification results are exactly given by Theorem 1, which in particular states that $q_\alpha$ and $q_{1-\alpha}$are unidentified. This lack of identifiability has been articulated by the modeling equivalence classes induced by the positive affine transformations we discussed in Section 3, and the identification results in Theorem 1 are fundamentally (sharp) set identification results as interpreted after Theorem 1 in Subsection 2.3.

Second, if $\phi$ is homogeneous of degree $p$, i.e., $\forall c > 0$, $\phi(ca, cb) = c^p \phi(a, b)$ for all $a, b \in \mathbb{R}$, then scale is not identifiable, and an interquantile-range scale normalization may again be imposed. Specifically, we may set

$$q_{1-\alpha} - q_\alpha = 1,$$

so that $q_{1-\alpha}$ can now be represented as known function of $q_\alpha$ by $q_{1-\alpha} = q_\alpha + 1$, and that only one "encryption key" $q_\alpha$ remains to be identified. If $\phi$ is nonlinear, we could potentially use the curvature of $\phi$ to identify $q_\alpha$. To illustrate this point more clearly, we consider in Subsection C.2 a concrete example in which $\phi$ is taken to be the cubic polynomial $\phi(a_1, a_2) = (a_1 + a_2)^3$, and show how to identify the location parameter that we can no longer normalize. Some of the methods presented in this example are more widely applicable, and will be used to discuss about more general specification of $\phi$ after Subsection C.2.

Third, we provide in Subsections C.3 and C.4 two sets of conditions under which location or scale is unidentifiable. Specifically, we propose two levels of generalization for "homogeneity", which underlies the scale normalization, and "translatability", which underlies the location normalization. Under each condition, we discuss what is identifiable and what is not, as well as the interpretation of the identification results.

## C.2 Illustrative Example: Cubic Coupling of Fixed Effects

We now consider a concrete example, where the fixed-effect coupling function $\phi$ is assumed to take the following cubic form:

$$\phi(a_1, a_2) := (a_1 + a_2)^3, \quad \forall a_1, a_2 \in \mathbb{R}.$$

We solve this example in detail to illustrate a more widely applicable strategy for identifying parameters that we can no longer normalize.

Clearly, as $\phi$ is homogeneous, we now may define a positive affine transformation that induces a new structure of modeling equivalence classes by the following:

$$\begin{aligned}
\hat{A}_i &:= c^{\frac{1}{3}} A_i, \\
\hat{w}(x_i, x_j) &:= c w(x_i, x_j) + b, \\
\hat{U}_{ij} &:= c U_{ij} + b.
\end{aligned}$$

It is straightforward to check that the above transformation indeed induces an equivalence relation, and we may as a result normalize

$$Q_\alpha [U_{ij}] = 0, \ Q_{1-a} [U_{ij}] = 1,$$

but leave as unknown

$$\theta_0 \equiv w(\overline{x}, \overline{x}), \ \forall \overline{x} \in Supp(X_i).$$

This is obviously equivalent to the alternative normalization of $\theta_0 = 0$ and $Q_{1-a}[U_{ij}] - Q_\alpha[U_{ij}] = 1$ while leaving $Q_\alpha[U_{ij}] = q_\alpha$ to be unknown, which is more in line with the general arguments laid out above in Subsection C.1. Yet for notational simplicity, we proceed with the normalization that leaves $\theta_0$ as unknown, and emphasize that nothing essential is different between the two approaches for normalization.

Given $\overline{\phi}(a) := \phi(a, a) = 8a^3$ and $\overline{\phi}^{-1}(z) = \frac{1}{2}z^{\frac{1}{3}}$, it becomes clear that the initial regions of identification of $F$ and $A$ are given by

$$\mathcal{F}_0 = \{0, 1\}, \ \mathcal{A}_0 = \left\{ -\frac{1}{2}\theta_0^{\frac{1}{3}}, \frac{1}{2}(1 - \theta_0)^{\frac{1}{3}} \right\},$$

with the "encryption key" $\theta_0$ being unknown. The in-fill and out-expansion arguments may then be deployed to extend the identification results to the whole real line, so we only need to establish identification of $\theta_0$.

We may identify $\theta_0$ by the curvature of$\phi$ via a feature of the first-order derivative of the induced function $\overline{\phi}(a) = 8a^3$: specifically, $\overline{\phi}'(a) = 0$ if and only if $a = 0$. To establish the identification result, we first *conjecture* that

$$\theta_0 = a_0$$

for an arbitrarily fixed (known) number $a_0 \in \mathbb{R}$, and then determine whether this conjecture can be falsified, i.e., whether $\mathbf{1}\{\theta_0 = a_0\}$ can be identified.

Given the conjecture $\theta_0 = a_0$, we can then "decrypt" the real line using $a_0$, as now the initial region of identification

$$\mathcal{A}_0 = \left\{ -\frac{1}{2}a_0^{\frac{1}{3}}, \frac{1}{2}(1 - a_0)^{\frac{1}{3}} \right\}$$

becomes effectively known. As a result, for any fixed number $a \in \mathbb{R}$, $F(a)$ and $\mathbf{1}\{A_i = a\}$ may be treated as known by our previous identification results. Now, we consider the following identification equation based on our conjecture $\theta_0 = a_0$:

$$\begin{aligned} & \frac{\partial}{\partial a}\mathbb{E}_{a_0}\left[ D_{ij} \middle| x_i = x_j, \ A_i = A_j = a \right] \\ =& \left(F \circ \overline{\phi}\right)'(a) \ \text{ if } \theta_0 = a_0 \\ =& F'\left(8a^3\right) \cdot 24a^2 \\ =& 0 \ \text{ if and only if } \ a = 0. \end{aligned} \tag{26}$$

provided that $F$ is differentiable at 0.

The first equality in (26) is an identification relationship, where the notation $\mathbb{E}_{a_0}[\cdot]$ emphasizes that the derivative $\frac{\partial}{\partial a}\mathbb{E}_{a_0}\left[ D_{ij} \middle| x_i = x_j, \ A_i = A_j = a \right]$ is an empirically observable one given the conjecture that $\theta_0 = a_0$, because the derivative, defined as the limit of ratios, can only be computed from data given a fixed and known $a_0$. Otherwise, the real line is

"encrypted" by an unknown $\theta_0$ so that we do not have an explicit scale to compute ratios in the form of

$$\frac{\mathbb{E}\left[D_{ij}\middle|\, x_i = x_j,\ A_i = A_j = a_1\right] - \mathbb{E}\left[D_{ij}\middle|\, x_i = x_j,\ A_i = A_j = a_2\right]}{a_1 - a_2}$$

when $a_1, a_2$ are potentially complicated functions of some unknown number $\theta_0$. However, given the conjecture $\theta_0 = a_0$, any two $a_1$ and $a_2$ can then be treated as effectively known numbers (as they are now known functions of the known number $a_0$) in the conditional expectations, so that both the conditional expectations and the differences between $a_1, a_2$, consequently their ratios in the above form along with the limit superior and limit inferior as $a_1 - a_2$ shrink to zero, are identified from data.

The second and the third equalities in (26), on the other hand, are completely theoretical, where the notation $F$ as before simply means the true unknown $F$. They, together with the first equality, provide a venue for falsification. Specifically, observe that

$$\frac{\partial}{\partial a}\mathbb{E}_{a_0}\left[D_{ij}\middle|\, x_i = x_j,\ A_i = A_j = a\right]\Big|_{a=0} = 0 \text{ if and only if } \theta_0 = a_0,$$

i.e., the empirical derivative corresponding to $\left(F \circ \overline{\phi}\right)'(0)$ is zero if and only if the empirical derivative is calculated under the correct conjecture. The intuition lies in that if $\theta_0 \neq a_0$, then the calculated empirical derivative

$$\frac{\partial}{\partial a}\mathbb{E}_{a_0}\left[D_{ij}\middle|\, x_i = x_j,\ A_i = A_j = a\right]\Big|_{a=0}$$

does not reproduce the true $\left(F \circ \overline{\phi}\right)'(0)$ but reproduces $\left(F \circ \overline{\phi}\right)'(b) \neq 0$ at some $b \neq 0$. The reason of this lies in the observation that each different $\theta_0$ "encrypts" the whole real line differently. More precisely, if $\theta_0 > a_0$, then the two initial points of identification

$$-\frac{1}{2}a_0^{\frac{1}{3}} > -\frac{1}{2}\theta_0^{\frac{1}{3}},\ \frac{1}{2}\left(1 - a_0\right)^{\frac{1}{3}} > \frac{1}{2}\left(1 - \theta_0\right)^{\frac{1}{3}}$$

shift monotonically, and as the ordering of individuals fixed effects $A_i$ is directly identifiable (as in 4.1) and thus invariant under arbitrary conjecture of $a_0$, the conceived location of "0" under the conjecture $a_0$ also shifts monotonically. It then follows that the conceived "0" coincides with the true 0 if and only if the conjecture $a_0$ is correct, i.e., $a_0 = \theta_0$.

We have thus established the identification of $\mathbf{1}\left\{\theta_0 = a_0\right\}$ for all $a_0 \in \mathbb{R}$, which implies that we have identified $\theta_0$. In view of the previous identification arguments and the scale normalization we have imposed, we summarize the identification of this example by the

following proposition:

**Proposition 3** (Nonparametric Identification with Cubic Coupling of Fixed Effects)**.** *Consider the nonparametric model (9) with cubic coupling of fixed effects:* $\phi(a_1, a_2) := (a_1 + a_2)^3$ *for all* $a_1, a_2 \in \mathbb{R}$*. Under Assumptions 1, 2, 3, 4 and 6, along with a further assumption that* $F$ *is differentiable at* $0$*, the modeling equivalence class of the unknown*

$$\left\{\left(cw + b,\ c^{\frac{1}{3}}A,\ cF^{-1} + b\right)\right\}$$

*to which the true value belong is identified, and no proper nonempty subset of it can be further identified.*

Essentially, the homogeneity of $\phi$ implies the lack of identification with respect to the scale parameter, which allows us to normalize the scale, while the nonlinearity of $\phi$, specifically via the uniqueness of "0" in terms of the first-order derivative of $\phi$, provides the identifying information for the location parameter. In this sense, Proposition 3 provides a sharper characterization of the identified set than Theorem 1, as the identified set in Proposition 3 features two degrees of freedom instead of three as in Theorem 1.

*Remark* 3 (**Identification without Assumption 1**). So far Assumption 1 has been maintained, requiring that the homophily function $w$ degenerates on the hyper-diagonal, i.e., $w(\overline{x}, \overline{x}) \equiv \theta \in \mathbb{R}$ for all $\overline{x} \in Supp(X_i)$. We now discuss why Assumption 1 is in fact no longer needed when $\phi$ is cubic, $\phi(a_1, a_2) = (a_1 + a_2)^3$, or more generally when the location of the homophily effect function $w$ can be identified via the curvature of $\phi$.

The identification arguments in the current Subsection start with the two-quantile normalization $Q_\alpha[U_{ij}] = 0$ and $Q_{1-a}[U_{ij}] = 1$, but leave $\theta_0 \equiv w(\overline{x}, \overline{x})$ as an unknown parameter. Without Assumption 1, we may as well write $\theta_{\overline{x}} := w(\overline{x}, \overline{x})$ for each value of on-support observable characteristics $\overline{x} \in Supp(X_i)$. The identification arguments in this Subsection, based on those in Subsections 4.1-4.3, are solely based on considerations of individuals that share the same observable characteristics. Hence, they apply to each particular $\overline{x} \in Supp(X_i)$, leading to the identification of $\theta_{\overline{x}}$ using the curvature of $\phi$. Carrying out this identification result for every $\overline{x}$, we have thus established the identification of $w$ on the whole hyper-diagonal, i.e., the value of $w(\overline{x}, \overline{x})$ for each $\overline{x}$, up to the two-quantile normalization we made.

We now explain the underlying intuition why Assumption 1 is no longer necessary. Essentially, Assumption 1 serves to "chain together" the identification results established for different observable characteristics $\overline{x}$ so that we can combine them in a meaningful way to identify the homophily effect function. Particularly, in Subsection 4.4, we need to consider individuals with different observable characteristics $x_i = \overline{x} \neq x_j = \overline{x}'$ but the same fixed effects $A_i = 0 = A_j$, which is sensible only if the meaning of "0" in the identification results for $\overline{x}$ coincides exactly with that for $\overline{x}'$, so that "$A_i = 0$" and "$A_j = 0$" can be combined as "$A_i = A_j = 0$". Assumption 1 achieves this alignment by assuming that the homophily effect function has no variation at all along the hyper-diagonal, which together with the location normalization on the distribution of $U_{ij}$ fix the location, or level, of the homophily effect function $w$ across its whole domain.

It is then clear why the cubic coupling function $\phi(a_1, a_2) = (a_1 + a_2)^3$ may substitute Assumption 1 for the role it played in the identification analysis. The fact that $\phi'(a) = 0$ if and only if $a = 0$ endows "0" with a unique defining property on the real line that remains invariant across different values of $\overline{x}$. This helps identify the location of $w$ that is consistent across all values of $\overline{x}$, automatically rendering Assumption 1 unnecessary. In addition, the whole function $w$ is now identifiable given the two-quantile normalization on $U_{ij}$.

More generally, other functional forms of $\phi$ may also substitute for Assumption 1, as long as the curvature of $\phi$ is rich enough to identify the location of $w$. This illustrates a form of substitutability between assumptions on the homophily effect function $w$ and assumptions on the coupling function $\phi$ in terms of their impacts on the identification power of the whole model.

## C.3 Generalized Homogeneity and Generalized Translatability

The example in the previous subsection considers a special case where the coupling function $\phi$ takes a cubic form, but the insights it provides on how to exploit any known curvatures of $\phi$ are more widely applicable, which is consistent with familiar idea that curvatures of a known function might be used for identification. In this Subsection we characterize more general conditions to which the identification arguments in the previous subsection can be adapted.

We say that $\phi$ is *generalized homogeneous* if, for any $c > 0$, there exists some (known) bijective function $g_c$ such that for all $(a_1, a_2) \in \mathbb{R}^2$,

$$\phi(g_c(a_1), g_c(a_2)) = c\phi(a_1, a_2). \tag{27}$$

This notion of homogeneity includes the usual homogeneity of order $p$ as as a special case.[13]

[13] If $\phi$ is homogeneous of order $p$, then $\phi(ca_1, ca_2) = c^p \phi(a_1, a_2)$ for all $a_1, a_2 \in \mathbb{R}$ and all $c > 0$. Then, for any given $c > 0$, setting $g_c(a) := c^{\frac{1}{p}} a$, we have $\phi(g_c(a_1), g_c(a_2)) = \phi\left(c^{\frac{1}{p}} a_1, c^{\frac{1}{p}} a_2\right) = c\phi(a_1, a_2)$ for all $(a_1, a_2)$, so $\phi$ is also generalized homogeneous.

If $\phi$ is generalized homogeneous, the following transformation will preserve modeling equivalence:

$$\hat{A}_i = g_{c^{1/p}}(A_i), \quad \hat{w} := cw + b, \quad \hat{U}_{ij} = cU_{ij} + b.$$

In particular, as $g_{c^{1/p}}$ is also bijective, Assumption 4 (conditional full support of fixed effect) holds for $\hat{A}_i$ if and only if it holds for $A_i$. Also, observational equivalence in terms of equation (9) is preserved, as

$$\phi\left(\hat{A}_i, \hat{A}_j\right) = \phi\left(g_c(A_i), g_c(A_j)\right) = c\phi(A_i, A_j). \tag{28}$$

In addition, the ordinal structure of fixed effects is also preserved, i.e., $\hat{A}_i > \hat{A}_j$ if and only if $A_i > A_j$ by equation (28), so that Assumption 6 ($\phi$ is strictly increasing in both arguments) is maintained, providing a consistent interpretation of $\hat{A}_i$ and $A_i$ both as individual degree heterogeneity. Given the above modeling equivalent transformation, we may impose the scale normalization $Q_{1-\alpha}[U_{ij}] - Q_\alpha[U_{ij}] = 1$ without loss of generality, and we are left effectively with one unknown location parameter, say $Q_\alpha[U_{ij}]$, to identify.[14]

Correspondingly, we now define a condition on $\phi$ that would allow the location normalization to be imposed. Specifically, we say that $\phi$ is *generalized translatable* if, $\forall b \in \mathbb{R}$, there exists a bijection $h_b : \mathbb{R} \to \mathbb{R}$ such that

$$\phi(h_b(a_1), h_b(a_2)) = \phi(a_1, a_2) + b, \ \forall a_1, a_2 \in \mathbb{R}. \tag{29}$$

For example, $\phi(a_1, a_2) = 1 + a_1^3 + a_2^3$ is both nonlinear and generalized translatable.[15] When $\phi$ is generalized translatable, the following transformation will preserve modeling equivalence:

$$\hat{A}_i = h_{2a}(A_i), \quad \hat{w} := w + b, \quad \hat{U}_{ij} = U_{ij} + 2a + b.$$

In particular, notice that $\phi\left(\hat{A}_i, \hat{A}_j\right) = \phi(A_i, A_j) + 2a$. Hence, we may now normalize a location parameter $Q_\alpha[U_{ij}]$ in addition to the location normalization $\theta_0 \equiv w(x_i, x_i) = 0$, so that again we are effectively left with one more parameter (a scale parameter) to identify. Clearly, if $\phi$ is both generalized homogeneous and generalized translatable, say, $\phi(a_1, a_2) = a_1^3 + a_2^3$,[16] then clearly $q_\alpha, q_{1-\alpha}$ are both unidentified, just as in the linear case.

[14] Recall that the interpretation here is not that we have identified the scale parameter by normalization, but that we have established the unidentifiability of the scale parameter by normalization and that we have characterized the underlying modeling equivalence relation among points that cannot be identified from each other.

[15] $\forall b \in \mathbb{R}$, setting $h_b(a) := \left(a^3 + \frac{1}{2}b\right)^3$, we then have (29).

[16] In this case, by considering $\tilde{A}_i = A_i^3$, one could argue that $\tilde{A}_i$ is an equivalent reformulation of the fixed

We summarize the results in this Subsection in the following lemma.

**Lemma (Identification with Potentially Nonlinear Coupling of Fixed Effects). 1**

*(i) If $\phi$ is generalized homogeneous, then the scale of the distribution of $U_{ij}$ (for instance, the interquantile range $Q_{1-\alpha}[U_{ij}] - Q_{\alpha}[U_{ij}]$) is unidentified.*

*(ii) If $\phi$ is generalized translatable, then the location of the distribution of $U_{ij}$ (for instance, its $\alpha$-th quantile $Q_{\alpha}[U_{ij}]$), is unidentified.*

Note that, in either of the cases in Lemma C.3, generalized homogeneity or generalized translatability helps characterize the identified set via the corresponding induced modeling equivalence classes, which in turn helps characterize the identified invariant properties within each modeling equivalence class. If both generalized homogeneity and generalized translatability are satisfied, then the characterization of the identified sets as in Theorem 1 essentially applies (with some minor corresponding adaptions).

## C.4 Periodic Homogeneity and Periodic Translatability

It should be pointed out, though, that generalized homogeneity or generalized translatability are sufficient, but not necessary, for the lack of point identification. For example, generalized translatability is stated for each translation constant $b \in \mathbb{R}$, so that each of the induced modeling equivalence class contains parameters are generalized translation of each other with any translation constant $b \in \mathbb{R}$. A weaker version of this condition, which will still implies the lack of point identification and induces a corresponding modeling equivalent class, is given by "periodic translatability" as defined below.

Specifically, we say that $\phi$ is *periodic translatable with period* $T > 0$ if there exists a bijective function $h_T : \mathbb{R} \to \mathbb{R}$ such that

$$\phi\left(h_T\left(a_1\right), h_T\left(a_2\right)\right) = \phi\left(a_1, a_2\right) + T, \quad \forall a_1, a_2 \in \mathbb{R}.$$

For example, $\phi\left(a_1, a_2\right) := \frac{1}{2}\left(a_1 + a_2 - \sin a_1 - \sin a_2\right)$ is periodic translatable[17] with period $T = 2\pi$ and a translation function $h_T(a) := a + T$, and in this case, it is intuitively clear that location of the distribution of $U_{ij}$ cannot be identified, as parameters that are periodic translations of each other with a period of $T = 2\pi$ under $h_T$ cannot be differentiated and must lie in the same modeling equivalence class.

effects under which the coupled effect $\tilde{A}_i + \tilde{A}_j$ is effectively reduced back to the linear case.

[17]Note that $\phi$ satisfies Assumption 6: it is symmetric, differentiable, strictly increasing and surjective.

Clearly, if $\phi$ is periodic translatable with every possible $T > 0$, then $\phi$ is generalized translatable. If $\phi$ is periodic translatable with a particular $T > 0$, then $\phi$ is periodic translatable with period $nT$ for every integer $n > 0$. If $\phi$ is periodic translatable with a particular $T > 0$ and $\phi$ is not periodic translatable with any period $t \in (0, T)$,[18] then the "phase", a description of the relative location within a period, becomes an invariant property of the modeling equivalence class induced by the periodic translation with period $T$, and thus may be identifiable under further assumption on the structure of $\phi$.

A similar weakened concept can be defined for generalized homogeneity: we say that $\phi$ is *periodic homogeneous with period* $T > 1$ if there exists a bijective function $g_T$ such that

$$\phi\left(g_T\left(a_1\right), g_T\left(a_2\right)\right) = T\phi\left(a_1, a_2\right), \quad \forall a_1, a_2 \in \mathbb{R}.$$

For a concrete example of $\phi$ that is periodic homogeneous but not generalized homogeneous, see Appendix E.

Are periodic homogeneity and periodic translatability not only sufficient but also necessary for the lack of point identification? We do not have a definite answer, though intuition suggests that, if there is no period $T$ under which the whole function $\phi$ is scalable or translatable, then every point on $\mathbb{R}$, the domain of $\phi$, should be endowed with some unique properties that provide identifiable information. We leave to future research a further study of this.

# D Unknown Coupling of Fixed Effects

As a further extension, we now consider a model with unknown coupling of fixed effects:

$$D_{ij} = \left\{w\left(X_i, X_j\right) + \phi\left(A_i, A_j\right) \geq U_{ij}\right\} \tag{30}$$

where $\phi : \mathbb{R}^2 \to \mathbb{R}$ is an *unknown* function that satisfies Assumption 6, i.e., $\phi$ is continuous, differentiable, strictly increasing and surjective in each of its arguments. For simplicity, assume that $A$ are independent of $X$ and that $A_i$ is continuously distributed on the whole real line $\mathbb{R}$. Then, take any strictly positive scalar $a \in \mathbb{R}_{++}$, any scalars $b, c \in \mathbb{R}$ and any continuous, differentiable, strictly increasing and surjective function $f : \mathbb{R} \to \mathbb{R}$, and define

$$\begin{aligned} \hat{A}_i &:= f\left(A_i, \right), \\ \hat{\phi}\left(\hat{A}_i, \hat{A}_j\right) &:= c\phi\left(f^{-1}\left(\hat{A}_i\right), f^{-1}\left(\hat{A}_j\right)\right) + a \end{aligned} \tag{31}$$

[18] For example, $\phi\left(a_1, a_2\right) := \frac{1}{2}\left(a_1 + a_2 - \sin a_1 - \sin a_2\right)$ is not periodic translatable with any period $t \in (0, 2\pi)$.

$$\hat{w}(x_i, x_j) := cw(x_i, x_j) + b,$$
$$\hat{U}_{ij} := cU_{ij} + a + b.$$

The above transformation of the unknown $(w, \phi, A, U)$ again maintains modeling equivalence. In particular, $\hat{\phi}$ remains symmetric, continuous, differentiable, strictly increasing and surjective, so that Assumption 6 is preserved. Hence, we may without loss of generality impose the following normalization:

$$\begin{aligned} \theta \equiv w(x_i, x_i) &= 0, \\ Q_\alpha [U_{ij}] &= 0, \\ Q_{1-\alpha} [U_{ij}] &= 1, \text{ for some } \alpha \in \left(0, \frac{1}{2}\right), \\ \overline{\hat{\phi}}(a) := \hat{\phi}(a, a) &= 2a \text{ for all } a \in \mathbb{R}. \end{aligned} \tag{32}$$

To obtain the last normalization on $\hat{\phi}$ in particular, we may take $f(a) := \frac{1}{2}\overline{\phi}(a) := \frac{1}{2}\phi(a, a)$. Then, as

$$f\left(\overline{\phi}^{-1}(2a)\right) = \frac{1}{2}\overline{\phi}\left(\overline{\phi}^{-1}(2a)\right) = \frac{1}{2} \cdot 2a = f\left(f^{-1}(a)\right).$$

we can deduce that $f^{-1}(a) = \overline{\phi}^{-1}(2a)$ and hence

$$\hat{\phi}(a, a) = \overline{\phi}\left(f^{-1}(a)\right) = \overline{\phi}\left(\overline{\phi}^{-1}(2a)\right) = 2a.$$

As a result of the normalization above, we now may start our identification arguments with all the normalization imposed in Section 3. Moreover, when individuals share the same fixed effects, the last normalization on $\phi$ take us essentially back to the case with linearly additive fixed effects. We now discuss how the identification arguments in Section 4 can be adapted to this setting.

Clearly, the identification arguments in Subsections 4.1 (identification of pairs of individuals with the same fixed effects) and 4.2 (identification of individuals with certain fixed effects) can be carried over to this case without changes. In particular, we may initially identify individuals with two levels of fixed effects, i.e., we can identify $\mathbf{1}\{A_i = 0\}$ and $\mathbf{1}\left\{A_i = \frac{1}{2}\right\}$, exactly as in Subsection 4.2.

For Subsection 4.3 (identification of the CDF $F$ and the fixed effects $A$), we might again adapt the "in-fill" and "out-expansion" arguments. Specifically, for the in-fill algorithm, we consider recursive iterations of the following identification equations:

$$F\left(\phi(a_i, a_j)\right) = \mathbb{E}\left[D_{ij} \mid x_i = x_j = \overline{x},\ A_i = a_i, A_j = a_j\right], \tag{33}$$

$$\mathbf{1}\left\{A_h = A_k = \frac{1}{2}\left(\phi\left(a_i, a_j\right)\right)\right\} = \mathbf{1}\left\{\mathbb{E}\left[D_{hk} \middle| x_h = x_k = \overline{x},\ A_h = A_k\right] = F\left(\phi\left(a_i, a_j\right)\right)\right\}, \quad (34)$$

Equation (33) extends the current knowledge of individuals with fixed effects $A_i = a_i$ and $A_j = a_j$ to the identification of $F$ at a new point $\phi(a_i, a_j)$, which is unknown but nevertheless characterizable by the two initial normalized quantiles $q_\alpha = 0$ and $q_{1-\alpha} = 1$. For example, though we do not know the value of $\phi\left(0, \frac{1}{2}\right)$, we can nevertheless identify $F\left(\phi\left(0, \frac{1}{2}\right)\right)$. In other words, .the composite function $F \circ \phi$ is identified at $\left(0, \frac{1}{2}\right)$, as well as $\left(\frac{1}{2}, 0\right)$ by symmetry. Moreover, notice that we know, by Assumption 6, that

$$\phi(0,0) = 0 < \phi\left(0, \frac{1}{2}\right) < 1 = \phi(0,1),$$

so we are indeed "in-filling" the interval $[0,1]$ in the domain of $F$. Equation (34) is exactly the same as the one we use in Subsection 4.3. For example, after having identified $F\left(\phi\left(0, \frac{1}{2}\right)\right)$, we may identify $\mathbf{1}\left\{A_i = \frac{1}{2}\phi\left(0, \frac{1}{2}\right)\right\}$, observing that $\frac{1}{2}\phi\left(0, \frac{1}{2}\right)$ bisects the interval $\left[0, \phi\left(0, \frac{1}{2}\right)\right]$ and lies strictly within $\left(0, \frac{1}{2}\right)$ as $\phi\left(0, \frac{1}{2}\right) < 1$. Hence, the interval $\left[0, \frac{1}{2}\right]$ in the range of $A_i$ is also being in-filled. Inductively, this again leads to the identification of $F$ and $A$ on the whole real line $\mathbb{R}$.

However, it should be pointed out that the identification of $F$ and $A$ on $\mathbb{R}$ is again "encrypted", now by the unknown coupling function $\phi$. This is very similar to the corresponding problem we encounter in Subsection C.2, where we consider a known cubic coupling function and find that the real line is encrypted by a single unknown parameter $(\theta_0)$. The difference, however, is that now, the identification of $F$ and $A$ on $\mathbb{R}$ is encrypted by an infinite-dimensional nonparametric object $\phi$, which is unknown outside the $\mathbb{R}^2$-diagonal (as $\phi(a,a) = 2a$ is known).

Complicated as it may seem, the problem we are now faced with may again be approached via the "conjecture and falsification" arguments as in Subsection C.2. Specifically, if we hypothesize an arbitrary specific form of $\phi$, subject to linear additivity on the $\mathbb{R}^2$-diagonal, we are then able to completely decrypt the identification results on $F$ and $A$ on the whole real line, leading to the identification of the nonparametric homophily effect $w$. Then we may try to falsify our conjectures by matching the properties of the conjectured $\phi$ (such as the curvatures of $\phi$ or $F \circ \phi$ outside the $\mathbb{R}^2$-diagonal) to the observable data. For example, given a conjectured $\phi$, $F$ and $A$ may be treated as known, then the theoretical and empirical derivatives, such as

$$\frac{\partial}{\partial a_1} F\left(\phi\left(a_1, a_2\right)\right) = \frac{\partial}{\partial a_1} \mathbb{E}_\phi\left[D_{ij} \middle| x_i = x_j, A_i = a_1, A_j = a_2\right]$$

$$= F^{'}\left(\phi\left(a_1,a_2\right)\right)\cdot\frac{\partial}{\partial a_1}\phi\left(a_1,a_2\right),$$

should be matched exactly.

As the space of admissible functions $\phi$ is infinite-dimensional, it seems hard to analytically or practically implement the identification strategy above. Yet the investigation and discussion of homogeneity and translatability in Subsections C.3 and C.4 suggest that we may obtain sharper characterization of the shape of the unknown coupling function $\phi$ within a potentially smaller space, if we impose some homogeneity and translatability restrictions on $\phi$. Homogeneity and translatability both impose restrictions on the structure of $\phi$ uniformly inside and outside the $\mathbb{R}^2$-diagonal, implying a form of uniformity on the whole domain of $\phi$ that may drastically reduce the space of admissible $\phi$. See the survey article by Matzkin (2007) for a related discussion on how homogeneity and additivity (essentially corresponding to "translatability" in our context) restrictions may help with nonparametric identification.

It should be pointed out, however, that even if $\phi$, after normalization, can be "identified" via the conjecture and falsification algorithm, the identified curvatures are all "measured" *relative to* the normalization that $\phi$ is additive on the $\mathbb{R}^2$-diagonal. However, we can never know whether or not the true $\phi$ is additive on the $\mathbb{R}^2$-diagonal, by the construction of the modeling equivalence transformation (31), which is no longer an affine transformation that preserves shapes. Being positive (strictly monotone) and continuous, however, transformation (31) preserves "lower-level" invariant properties such as orders and topological structures. Such invariant properties seem at least relevant for some *qualitative* results and counterfactuals.

# E An Example of Periodic Homogeneity

In this section, we provide an example of a nonlinear function $\phi$ that is periodic homogeneous but not generalized homogeneous.

We first construct two auxiliary functions $m(\cdot)$ and $\lambda(\cdot)$. Define

$$m(a) := \lfloor \log_2 |a| \rfloor \equiv \max\left\{m\in\mathbb{Z}:\ m\le\log_2|a|\right\},\quad \forall a\in\mathbb{R}\backslash\{0\}$$

and set

$$\lambda(a) := \begin{cases} a+\frac{a}{|a|}\cdot\frac{2^{m(a)}}{4\pi}\sin\left(\frac{4\pi}{2^{m(a)}}a\right), & \text{for } a\in\mathbb{R}\backslash\{0\},\\ 0, & \text{for } a=0.\end{cases}$$

Figure 1 plots the function $\lambda$ around zero.

First, we show that $\lambda$ is continuous, differentiable and strictly increasing on $\mathbb{R}$. Clearly,

Figure 1: Plot of function $\lambda(a)$

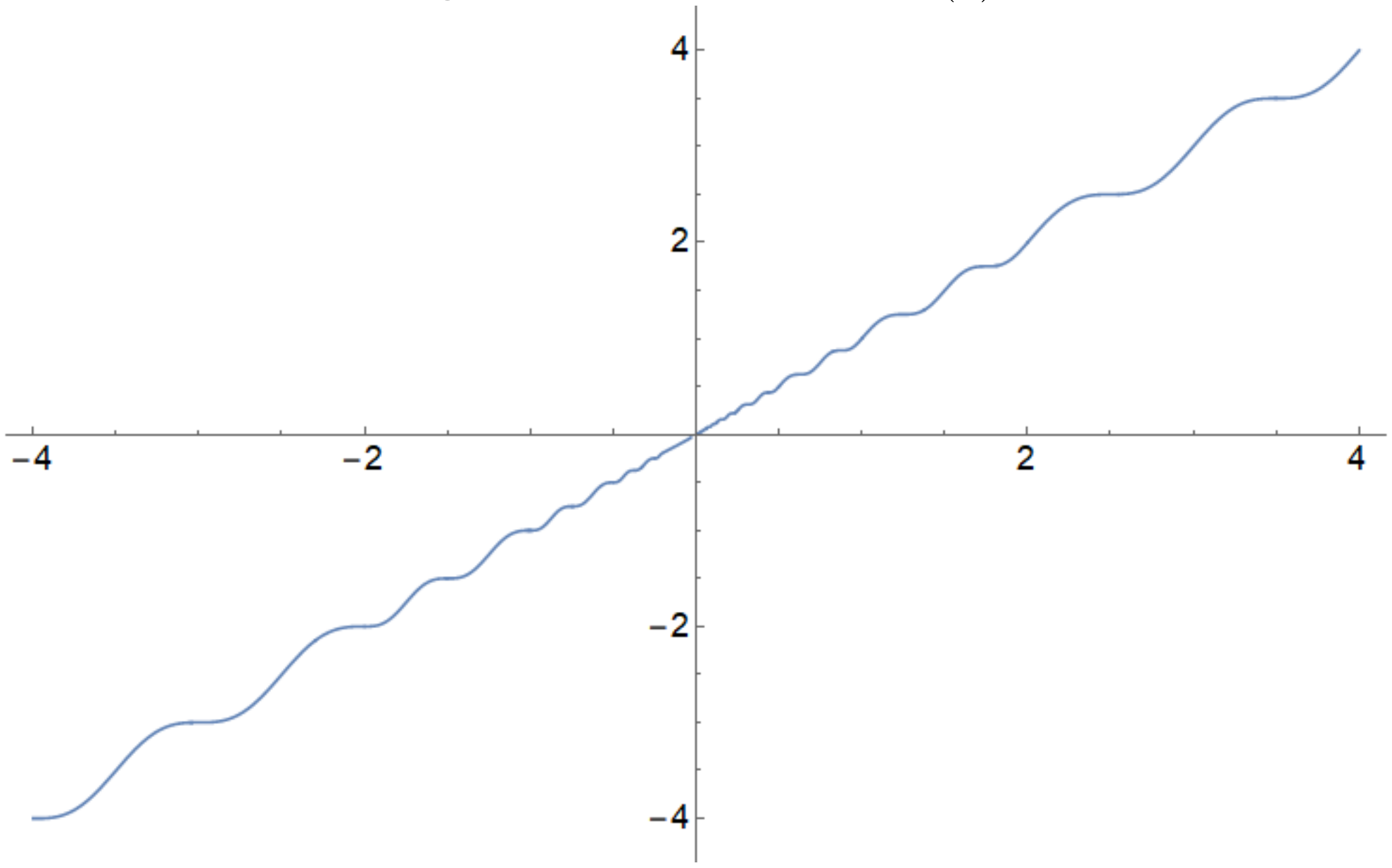

$\lambda$ is continuous, differentiable and strictly increasing on $(2^m, 2^{m+1})$ for every $m \in \mathbb{N}$:

$$\lambda(a) = a + \frac{2^m}{4\pi} \sin\left(\frac{4\pi}{2^m} a\right)$$
$$\lambda'(a) = 1 + \cos\left(\frac{4\pi}{2^m} a\right) \geq 0$$

Moreover, $\lambda$ is also differentiable at each boundary point $2^m$ with $\lambda'(2^m) = 2$, so $\lambda$ is strictly increasing and differentiable on $(0, \infty)$. As $\lambda$ is an odd function, i.e., $\lambda(a) = -\lambda(-a)$, $\lambda$ is also differentiable and strictly increasing on $(-\infty, 0)$. Finally, $\lambda$ is also differentiable at $a = 0$, because

$$\begin{aligned}\left|\frac{\lambda(a) - \lambda(0)}{a}\right| &= \left|1 + \frac{1}{4\pi} \cdot \frac{2^{m(a)}}{|a|} \sin\left(\frac{4\pi}{2^{m(a)}} a\right)\right| \\ &\leq 1 + \frac{1}{4\pi} \cdot \frac{2^{m(a)}}{|a|} \\ &\to 1 + 0 \text{ as } m(a) \to -\infty \text{ when } |a| \to 0.\end{aligned}$$

Thus $\lambda'(0) = 1$. In summary, $\lambda$ is differentiable[19] and strictly increasing on $\mathbb{R}$.

Second, we show that $\lambda$ is periodic homogeneous function with period $T = 2$. For $a = 0$

[19] Note, however, that $\lambda$ is not continuously differentiable, as $\lambda'$ is not continuous at 0.

this is trivial. For $a \neq 0$,

$$\begin{aligned}\lambda(2a) &= 2a + \frac{2a}{|2a|} \cdot \frac{2^{m(2a)}}{4\pi} \sin\left(\frac{4\pi}{2^{m(2a)}} 2a\right) \\ &= 2a + \frac{2a}{|2a|} \cdot \frac{2^{m(a)+1}}{4\pi} \sin\left(\frac{4\pi}{2^{m(a)+1}} 2a\right) \\ &= 2\left[a + \frac{a}{|a|} \cdot \frac{2^{m(a)}}{4\pi} \sin\left(\frac{4\pi}{2^{m(2a)}} 2a\right)\right] \\ &= 2\lambda(a),\end{aligned}$$

as the observation that

$$\begin{aligned}m(2a) &= \max\{m \in \mathbb{Z}: \ m \leq \log_2 |2a|\} \\ &= \max\{m \in \mathbb{Z}: \ m \leq 1 + \log_2 |a|\} = m(a) + 1, \quad \forall a \neq 0.\end{aligned}$$

Now, define the fixed effect coupling function $\phi : \mathbb{R} \times \mathbb{R} \to \mathbb{R}$ by

$$\phi(a_1, a_2) := \lambda(a_1 + a_2), \quad \forall a_1, a_2 \in \mathbb{R}.$$

Clearly, $\phi$ satisfies Assumption 6: it is symmetric, differentiable, strictly increasing and surjective in each of its arguments.

Clearly, $\phi$ is periodic homogeneous with period $T = 2$:

$$\begin{aligned}\phi(2a_1, 2a_2) &= \lambda(2(a_1 + a_2)), \\ &= 2\lambda(a_1 + a_2) = 2\phi(a_1, a_2), \quad \forall a_1, a_2 \in \mathbb{R}.\end{aligned}$$

However, $\phi$ is not generalized homogeneous: for $c = 3$, there does not exist any function $g : \mathbb{R} \to \mathbb{R}$ such that

$$\phi(g(a_1), g(a_2)) = 3 \cdot \phi(a_1, a_2), \quad \forall a_1, a_2 \in \mathbb{R}.$$

To see this, notice that $\phi(0.5, 0.5) = 1$ and $\phi(1.5, 1.5) = 3$ imply

$$g(0.5) = 1.5.$$

In the meanwhile, $\phi(0.9, 0.9) \approx 1.7532$ and $\phi(2.7851, 2.7851) \approx 3 \times \phi(0.9, 0.9)$ imply that

$$g(0.9) \approx 2.7851.$$

However,

$$\phi(1.5, 2.7851) \approx 4.5335 \neq 3.9730 \approx 3 \cdot \phi(0.5, 0.9),$$

so $\phi$ is not generalized homogeneous.

# F Bounded Conditional Support of Fixed Effects

We now investigate the case where Assumption 4 (Conditional Full Support of Fixed Effects) is relaxed to the following version:

**Assumption. 4"** *(Bounded Conditional Support of Fixed Effects)* $\forall x \in Supp(X_i), Supp(A_i | X_i = x) \equiv [\underline{a}, \overline{a}]$ *for some* $-\infty < \underline{a} < \overline{a} < \infty$.

We focus on the identification of model (3) with Assumptions 1, 2 and 3 unchanged. We impose the following normalization: for some fixed $0 < \underline{\alpha} < \overline{\alpha} < 1$, set

$$w(\overline{x}, \overline{x}) = 0, \; F^{-1}(\underline{\alpha}) = 0, \; F^{-1}(\overline{\alpha}) = 1,$$

which differs almost trivially from the normalization imposed in Section 3.

We now proceed to establish identification of $w, A, F^{-1}$ on its whole support.

Following Subsection 4.1, we may still identify a complete ordering of fixed effects among individuals that share the same observable characteristics $\overline{x}$.

Following Subsection 4.2, we may still identify two initial levels of fixed effects, $\mathbf{1}\{A_i = 0\}$ and $\mathbf{1}\left\{A_i = \frac{1}{2}\right\}$, provided that

$$F(2\underline{a}) \leq \underline{\alpha} < \overline{\alpha} \leq F(2\overline{a}).$$

As $\underline{\alpha}, \overline{\alpha}$ can be arbitrarily chosen, the above can always be guaranteed. Though the normalization may induce a change in the nominal values of the boundary points of $Supp(A_i | X_i = x)$ relative to the underlying normalization, we slightly abuse notations and write $Supp(A_i | X_i = x) = [\underline{a}, \overline{a}]$ under any imposed normalization, as the normalization does not change any topological structure so that the representation of $Supp(A_i | X_i = x)$ by $[\underline{a}, \overline{a}]$ may be regarded as trivial relabeling of $\underline{a}$ and $\overline{a}$. Then, by monotonicity we have

$$\underline{a} < 0 < \overline{a}.$$

Following Subsection 4.3, we may carry out the "in-fill" algorithm exactly as before and identify $F$ on $[0, 1]$ and $A$ on $\left[0, \frac{1}{2}\right]$.

 The "out-expansion" algorithm can also be carried out exactly as before, but the identification of $A_i$ must stop at

$$\overline{\mathcal{A}} = [\underline{a}, \overline{a}] \subsetneq \mathbb{R}.$$

As the identification of $F$ relies recursively on the identification of $A_i$, the identification of $F$ stops at

$$\overline{\mathcal{F}}_0 = [2\underline{a}, 2\overline{a}] \subsetneq \mathbb{R}$$

with a range space given by

$$Range\left(\overline{\mathcal{F}}\right) = [F(2\underline{a}), F(2\overline{a})].$$

This is the largest region of identification we can achieve with the "in-fill and out-expansion" algorithm. However, we can now exploit variations in $w(\overline{x}, \underline{x})$ in a recursive way, as explained below.

Following Subsection 4.4, we now may identify the nonparametric homophily effect function $w$ on the region where there exist

$$a \in [\underline{a}, \overline{a}], \quad b \in [F(2\underline{a}), F(2\overline{a})]$$

such that

$$F(w(\overline{x}, \underline{x}) + 2a) = b,$$

which gives us

$$w(\overline{x}, \underline{x}) = F^{-1}(b) + 2a.$$

Hence, $w$ is identified on the region

$$\overline{\mathcal{W}}_0 = \left\{ (\overline{x}, \underline{x}) \in Supp(X_i)^2 : \; w(\overline{x}, \underline{x}) = F^{-1}(b) + 2a : \begin{array}{c} a \in [\underline{a}, \overline{a}], \\ b \in [F(2\underline{a}), F(2\overline{a})] \end{array} \right\}$$

$$= \left\{ (\overline{x}, \underline{x}) \in Supp\,(X_i)^2 : w\,(\overline{x}, \underline{x}) \in [2\underline{a} + 2\underline{a}, 2\overline{a} + 2\overline{a}] \right\}$$
$$= \left\{ (\overline{x}, \underline{x}) \in Supp\,(X_i)^2 : w\,(\overline{x}, \underline{x}) \in [4\underline{a}, 4\overline{a}] \right\}.$$

We write the range space of $\overline{\mathcal{W}_0}$ under $w$ as

$$w\left(\overline{\mathcal{W}}_0\right) := Supp\,(w\,(X_i, X_j)) \bigcap [4\underline{a}, 4\overline{a}],$$

and define

$$\overline{w}_0 := \max w\left(\overline{\mathcal{W}}_0\right), \quad \underline{w}_0 := \min w\left(\overline{\mathcal{W}}_0\right),$$

which are well-defined as both $Supp\,(w\,(X_i, X_j))$ and $[4\underline{a}, 4\overline{a}]$ are closed sets in $\mathbb{R}$. Moreover, as $w\,(\overline{x}, \overline{x}) \equiv 0$, we have

$$\underline{w}_0 \leq 0 \leq \overline{w}_0.$$

We now carry out a recursive "out-expansion" algorithm using variations in observable characteristics $(\overline{x}, \underline{x})$ in addition to variations in fixed effects $A_i, A_j$. Consider pairs of individuals with observable characteristics $X_i = \overline{x}$ and $X_j = \underline{x}$ such that

$$w\,(\overline{x}, \underline{x}) \in c \in \{\overline{w}_0, \underline{w}_0\},$$

and individual fixed effects such that

$$A_i = a_i \in \overline{\mathcal{A}}, \quad A_j = a_j \in \overline{\mathcal{A}}.$$

We can then identify

$$F\,(c + a_i + a_j) = \mathbb{E}\left[ D_{ij} \middle| X_i = \overline{x},\ X_j = \underline{x},\ A_i = a_i,\ A_j = a_j \right],$$

extending the identification of $F$ to

$$\overline{\mathcal{F}}_1 := \left\{ c + a_i + a_j : \begin{array}{c} c \in \{\overline{w}_0, \underline{w}_0\} \\ a_i \in \overline{\mathcal{A}} = [\underline{a}, \overline{a}] \\ a_j \in \overline{\mathcal{A}} = [\underline{a}, \overline{a}] \end{array} \right\} = [\underline{w}_0 + 2\underline{a}, \overline{w}_0 + 2\overline{a}].$$

The extension of $\overline{\mathcal{F}}$ to $\overline{\mathcal{F}}_1$ is strict, i.e., $\overline{\mathcal{F}} \subsetneq \overline{\mathcal{F}}_1$, if and only if the following assumption holds:

**Assumption. 7-0** (Locally Nondegenerate Support of Homophily Effects):

$$\underline{w}_0 < 0 < \overline{w}_0.$$

As the fixed effects have already been fully identified on $\overline{\mathcal{A}}$, we now only need to recursively extend the identification of $w$. Now, with the knowledge of $F$ on $\overline{\mathcal{F}}_1$, we may identify $w$ on the region where there exist

$$a \in [\underline{a}, \overline{a}], \quad b \in F\left(\overline{\mathcal{F}}_1\right),$$

which establishes identification of $w$ on the following region

$$\begin{aligned}\overline{\mathcal{W}}_1 &:= \left\{ (\overline{x}, \underline{x}) \in Supp\left(X_i\right)^2 : w\left(\overline{x}, \underline{x}\right) = F^{-1}\left(b\right) + 2a, \begin{array}{c} a \in [\underline{a}, \overline{a}], \\ b \in F\left(\overline{\mathcal{F}}_1\right) \end{array} \right\} \\ &= \left\{ (\overline{x}, \underline{x}) \in Supp\left(X_i\right)^2 : w\left(\overline{x}, \underline{x}\right) \in [\underline{w}_0 + 4\underline{a}, \overline{w}_0 + 4\overline{a}] \right\}\end{aligned}$$

and the range space of $\overline{\mathcal{W}}_1$ under $w$ is given by

$$w\left(\overline{\mathcal{W}}_1\right) := Supp\left(w\left(X_i, X_j\right)\right) \bigcap [\underline{w}_0 + 4\underline{a}, \overline{w}_0 + 4\overline{a}].$$

Now define

$$\overline{w}_1 := \max w\left(\overline{\mathcal{W}}_1\right) \geq \overline{w}, \quad \underline{w}_1 := \min w\left(\overline{\mathcal{W}}_1\right) \leq \underline{w}.$$

The extension from $\overline{\mathcal{W}}$ to $\overline{\mathcal{W}}_1$ is nontrivial, i.e., $\overline{\mathcal{W}} \subsetneqq \overline{\mathcal{W}}_1$ and $w\left(\overline{\mathcal{W}}\right) \subsetneqq w\left(\overline{\mathcal{W}}_1\right)$, if and only if the following assumption holds:

**Assumption. 7-1** (1st-Order Recursive Nondegenerate Support of Homophily Effects):

$$\underline{w}_1 < \underline{w}_0, \quad \textit{and} \quad \overline{w}_0 < \overline{w}_1.$$

Continuing recursively, define

$$\begin{aligned}\overline{\mathcal{F}}_m &:= \left[\underline{w}_{m-1} + 2\underline{a}, \overline{w}_{m-1} + 2\overline{a}\right], \\ \overline{\mathcal{W}}_m &:= \left\{ (\overline{x}, \underline{x}) \in Supp\left(X_i\right)^2 : w\left(\overline{x}, \underline{x}\right) \in \left[\underline{w}_{m-1} + 4\underline{a}, \overline{w}_{m-1} + 4\overline{a}\right] \right\}, \\ \overline{w}_m &:= \max w\left(\overline{\mathcal{W}}_m\right) \geq \overline{w}_{m-1}, \\ \underline{w}_m &:= \min w\left(\overline{\mathcal{W}}_m\right) \leq \underline{w}_{m-1},\end{aligned}$$

and impose the following assumption:

**Assumption. 7-m** ($m$-th-Order Recurisve Nondegenerate Support of Homophily Effects):

$$\underline{w}_m < \underline{w}_{m-1}, \quad and \quad \overline{w}_{m-1} < \overline{w}_m.$$

We can now formally present our main result.

**Proposition 4** (Identification under Bounded Conditional Support of Fixed Effects)**.** *Suppose Assumption 1, 2, 3 and 4" hold. Moreover, suppose that Assumptions 7-m holds for all* $m = 0, 1, ..., M-1$. *Then, under the imposed normalization,* $A_i$ *is fully identified on* $\overline{\mathcal{A}} = [\underline{a}, \overline{a}]$, $w$ *is identified on* $\overline{\mathcal{W}}_M$ *and* $F$ *is identified on* $\overline{\mathcal{F}}_M$.

**Corollary 1.** *Suppose Assumption 1, 2, 3 and 4" hold.*

*(a) If* $Supp\left(w\left(X_i, X_j\right)\right) = \mathbb{R}$, *then* $A_i, w, F$ *are fully identified up to the imposed normalization.*

*(b) If there exist some* $M \in \mathbb{N} \cup \{\infty\}$ *such that*

$$Supp\left(w\left(X_i, X_j\right)\right) = w\left(\overline{\mathcal{W}}_M\right),$$

*then* $A_i$ *and* $w$ *are fully identified up to the imposed normalization, while* $F$ *is on* $\overline{\mathcal{F}}_M$.

In Corollary 1(a), the supposition $Supp\left(w\left(X_i, X_j\right)\right) = \mathbb{R}$ can be satisfied even in the presence of discrete characteristics, as long as there exists a dimension of continuous covariate with large variations that induces the full support of $w\left(X_i, X_j\right)$ a la Horowitz (1992).

In Corollary 1(b), it seems that the identification of $F$ is incomplete, as it is only identified on $\overline{\mathcal{F}}_M$, which may be strict subset of $\mathbb{R}$. However, noticing that all variations in both the observable characteristics $X_i, X_j$ and the unobserved individual fixed effects $A_i, A_j$ have been exhausted, we may no longer be interested in the values of $F$ outside $\overline{\mathcal{F}}_M$. For all practical purposes, we may redefine

$$\tilde{F}(a) := \begin{cases} 0, & a \in \left(-\infty, \min \overline{\mathcal{F}}_M\right), \\ F(a), & a \in \overline{\mathcal{F}}_M, \\ 1, & a \geq \left[\max \overline{\mathcal{F}}_M, \infty\right). \end{cases}$$

as this transformation, though it would violate modeling equivalence, preserves observational equivalence.